\documentclass[a4paper,fleqn,usenatbib]{mnras}

\usepackage[T1]{fontenc}
\usepackage{ae,aecompl}

\usepackage{graphicx}	
\usepackage{amsmath}	
\usepackage{amssymb}	

\title[CO/CN in galaxies with ALMA]{Exploring the CO/CN line ratio in
  nearby galaxies with the 
ALMA archive}

\author[C. D. Wilson et al.]{
Christine D. Wilson$^{1}$\thanks{E-mail: wilsoncd@mcmaster.ca}
\\
$^{1}$Department of Physics and Astronomy, McMaster University, 1280
Main St. W., Hamilton, Ontario L8S 4M1, Canada\\
}

\date{Accepted XXX. Received YYY; in original form ZZZ}

\pubyear{2017}

\begin{document}
\label{firstpage}
\pagerange{\pageref{firstpage}--\pageref{lastpage}}
\maketitle

\begin{abstract}
We describe an archival project using Cycle 0 data from the Atacama Large 
Millimeter/submilleter Array to survey the CO/CN line ratio in 17 nearby
galaxies. CN is an interesting molecule that traces dense gas exposed to
ultraviolet radiation and its $N=1-0$ lines can be observed simultaneously 
with the CO $J=1-0$ line. We identify 8 galaxies with distances $< 200$ Mpc 
for which both lines are detected.
Signal-to-noise matched CO/CN ratios range from as low as 7 to as 
high as 65, while ratios using the total detected flux 
range from 20 to 140. Spatial
variations greater than a factor of 3 are seen in several galaxies.
These 
line ratio changes are likely due to changes in the [CN]/[H$_2$]
abundance ratio and/or the CN
excitation.  Additional measurements of the warm gas pressure and 
the CN excitation should help to distinguish between these two possibilities.
3 of the 4 active galactic nuclei in our sample show CO/CN line ratios that are
roughly a factor of 2-3 larger than those seen in starburst-dominated
regions, which may be in conflict with models of molecular abundances
in X-ray dominated regions.

\end{abstract}

\begin{keywords}
galaxies: ISM -- galaxies: starburst -- ISM: molecules -- ISM:
abundances -- galaxies: abundances
\end{keywords}



\section{Introduction}

Molecular gas is an important component of the interstellar medium in
galaxies at both low and high redshift. As the fuel for star
formation, it provides a critical link with the stars that are the
bulk of the baryonic mass in individual galaxies. Because molecular
hydrogen itself is largely invisible at the low temperatures ($<50$ K)
where the majority of the molecular gas is found, carbon monoxide 
(CO) is typically used to measure  the molecular
gas. Other molecules such as HCN and HCO$^+$ are used to trace the higher
density gas that is more directly associated with recent star
formation \citep{gao2004,wu2010,kennicutt2012}.
However, most
molecular lines are
typically an order of magnitude fainter than  low-$J$ transitions
of CO and thus only global measurements made with single dish
telescopes are typically available for other molecular species
\citep[e.g.][]{aalto2002,gao2004,juneau2009}.

A particularly interesting  molecule is the cyanide radical (CN), which
can be produced
from HCN via photo-dissociation and from reactions involving C$^+$ and
C \citep[e.g.][]{boger2005}. CN thus provides a probe of dense gas
exposed to strong ultraviolet radiation fields in photon-dominated
regions (PDRs). Formation of CN at
high column densities can be driven by enhanced cosmic ray ionization rates
\citep[e.g.][]{boger2005} or by X-rays near an active galactic nucleus
(AGN) as shown by models of X-ray dominated regions (XDRs) 
\citep[e.g.][]{meijerink2007}.  
The CN $N=1-0$ transition consists of 9 hyperfine lines blended into two
groups, with the stronger group corresponding to the $J=3/2-1/2$
transitions and the weaker group the $J=1/2-1/2$ transitions.
A recent large-area (10x10 pc) map of the Orion B molecular cloud 
shows a strong correlation of the CN emission with the
ultraviolet radiation field \citep{gratier2017}. However, a
large (40x50) pc map of W51 by \citet{watanabe2017} shows significant
CN emission throughout 
the cloud with a relatively constant CN/$^{13}$CO line ratio.
The few extragalactic observations of CN $N=1-0$
prior to the Atacama Large 
Millimeter/submillimeter Array (ALMA) were primarily single-dish
observations of infrared 
luminous galaxies \citep{aalto2002}, nearby Seyfert
galaxies \citep{perez-beaupuits2007,perez-beaupuits2009,aladro2013}
and AGNs \citep{aalto2007,chung2011}.
\citet{fuente2005} detected CN emission at 3 locations in the nearby starburst
galaxy M82. Typical CO($J=1-0$)/CN($N=1-0;J=3/2-1/2$)
integrated intensity 
ratios are in the range of 10-20 in these systems. 

\pagebreak

A few of the brightest galaxies were mapped in CN at higher
angular resolution prior to ALMA. 
\citet{ginard2015} observed M82 at $\sim 2.5^{\prime\prime}$
resolution in several 
molecular lines, including CN $N=1-0$, with the Plateau de Bure
Interferometer (PdBI). They found significant
variations in the CN abundance within M82 as well as a
correlation of the CN/N$_2$H$^+$ line ratio with H41$\alpha$. These CN
variations 
can be explained in chemical models as due to changes in the strength
of the ultraviolet radiation field and the average size of the
molecular clouds.
\citet{garcia-burillo2010} mapped the nearby Seyfert 2 galaxy NGC 1068
in the CN N=2-1 line and the SiO J=2-1 line at
$1-3^{\prime\prime}$ resolution with the PdBI.
The data reveal a high CN abundance in the
circumnuclear disk that cannot be produced by either shock or
PDR models. \citet{garcia-burillo2010}
suggest that the CN and SiO abundances imply that the circumnuclear
disk region is a giant X-ray dominated region.

The fast mapping speed and high sensitivity of 
ALMA have opened the possibility to trace
emission from rarer molecules and isotopes across a wide variety of
galaxies and environments. ALMA's imaging capabilities also provide
the opportunity to study spatial variations in molecular emission
lines inside galaxies. In addition, ALMA's broad spectral coverage  provides
observations of a wide range of molecular
species in just a few observations \citep[e.g.][]{meier2015}.
These increased capabilities have produced a huge increase in the amount and
quality of data available for nearby galaxies for the 
$N=1-0$ transitions of CN, which lie
less than 2 GHz away from the  CO $J=1-0$ line. This
fortuitous arrangement means that CN $N=1-0$ can be easily observed whenever
the CO $J=1-0$ line is the primary target. Such serendipitous
observations of CN were particularly common during ALMA Cycle 0,
when only limited correlator modes were available. For
observations of nearby galaxies, the most commonly used correlator
mode in Cycle 0 covered both the CO and the CN ground state lines with
a spectral resolution of better than 1 km s$^{-1}$. Programs in later
cycles often observed CN along with the CO $J=1-0$ line, but
sometimes in ALMA's ``continuum'' mode with significantly lower
velocity resolution on the CN line.

During ALMA Cycle 0, CN was one of the $\sim 50$ molecular lines in the 3
mm window mapped  in the nearby 
starburst galaxy NGC 253 \citep{meier2015}. These data
show that the CN lines are optically thin and that the CO/CN line
ratio increases with radius.
Although the CN and HCN lines could not be compared
directly because the maps had different angular resolutions, the 
HCN/C$_2$H ratio shows a similar increase with radius. Since
both CN and C$_2$H are tracers of photon dominated regions (PDRs),
\citet{meier2015} concluded that PDRs account for a lower fraction of
the dense gas in the outer disk compared to the inner disk.
\citet{saito2015} also find spatially varying abundances of CN and
other molecules in the luminous infrared galaxy (LIRG) merger VV 114. Their
suggestion that emission from the overlap region is dominated by
shocks is supported by the detection of methanol from only the overlap
region \citep{saito2017}.
\citet{iono2013} argue that the eastern
nucleus of VV114 contains an AGN.

\citet{sakamoto2014} found faint, broad line wings in the CN emission
from the LIRG merger NGC 3256, which suggests that some
of the CN emission originates in nuclear outflows with 
an enhanced CN abundance.
They use the outflow properties to argue that the
southern nucleus hosts a (possibly dormant) AGN.
Observing in the 350~GHz window,
\citet{nakajima2015} find spatial variations in the $^{13}$CO/CN
ratio in NGC 1068, which they attribute to enhanced CN emission 
in the XDR of the circumnuclear
disk. However, they note that changes in the $^{13}$CO abundance
between the circumnuclear disk and the starburst ring may also play a
role. 
\citet{garcia-burillo2014} find dense gas
tracers such as HCN and HCO$^+$  in the
circumnuclear disk of NGC 1068 and also in the outflow. 
Line ratios relative to CO are enhanced by a factor of 10
compared to the gas in the 1.3 kpc starburst ring, which they
attribute to the UV/X-ray radiation coming from the AGN.
\citet{garcia-burillo2017} find
that PDR models can account for the aboundance of C$_2$H in the
starburst ring of NGC 1068, but that the outflow emission requires
time-dependent 
chemistry as well as shocks.

In this
paper, we present an initial survey of the CN $N=1-0;J=3-2/1-2$ line (the stronger
of the two 3 mm CN emission lines) 
observed simultaneously with the CO $J=1-0$ line in
17 nearby galaxies drawn from 6 separate Cycle 0 programs. 
\S\ref{data} describes
the selection of the CN galaxy sample, the
image processing, and the method used to measure the CO/CN ratios.  
\S\ref{discussion} discusses the properties of CN in the different galaxies
and presents possible explanations for the  observed
variations in the CO/CN line ratios. \S\ref{conclusions} summarizes
the conclusions and future work. Some details on the individual
galaxies in the sample are given in Appendix~\ref{individual-galaxies}
and upper limits for the galaxies without CN detections are given in
Appendix~\ref{app-upperlimits}. 


\section{Sample selection and analysis}
\label{data}

\begin{table}
	\centering
	\caption{ALMA Cycle 0 projects with CO and CN observations of
          nearby galaxies}
	\label{table-projects}
	\begin{tabular}{lll} 
		\hline
		Project Code & Galaxy & ALMA data reference\\
		\hline
2011.1.00099.S & 20 galaxies & \citet{ueda2014}\\
2011.1.00172.S  & NGC 253 & \citet{bolatto2013a}\\ 
2011.1.00467.S  & VV 114 & \citet{iono2013}\\
2011.1.00525.S  & NGC 3256 & \citet{sakamoto2014}\\
2011.1.00645.S  & NGC 1377 & ... \\
2011.1.00772.S  & M83 & \citet{freeman2017}\\ 
		\hline
	\end{tabular}
\end{table}

\begin{table*}
	\centering
	\caption{Galaxies in the ALMA Cycle 0 CN sample}
	\label{table-sample}
	\begin{tabular}{llcccccc} 
		\hline
		Galaxy & beam size, PA & beam & $\Delta V$ & CN
                sensitivity & CO sensitivity &    CN mom0 & CO mom0   \\
... & ($\prime\prime$, degrees) & (kpc) & (km s$^{-1}$) & (mJy beam$^{-1}$) &
             (mJy beam$^{-1}$) & cut ($\sigma$) & multiplier\footnotemark[1] \\
		\hline
AM 2055-425 & 1.5x1.2, 45 & 1.2x1.0 & 20 & 1.9 & 2.3 & 2 & 15  \\ 
AM 2246-490 & 1.45x1.30, 0 & 1.3x1.2 & 20 & 1.9 & 2.4 & 2 & 6 \\
NGC 3256 & 3.2x2.1, 5 & 0.67x0.44 & 20 & 1.0 & 1.3 & 3 & 20  \\ 
VV 114 & 5.5x3.1, -80 & 2.1x1.2 & 30 & 1.1 & 1.3 & 3 & 30 \\
AM 1300-233 & 2.5x1.3, -76 & 1.1x0.6 & 40 & 1.2 & 1.2 & 3 & 20  \\
NGC 253 & 4x3, -34 & 0.068x0.051 & 5 & 4.7 & 5.2 & 3 & 20  \\ 
M83 & 2.1x2.1, 0 & 0.052x0.052 & 10 & 4.0 & 6.9 & 2  & 20  \\ 
NGC 1377 & 1.7x1.25, -87 & 0.19x0.15 & 20 & 1.2 & 1.7 & 2 & 30  \\ 
		\hline
	\end{tabular}
\\
\footnotemark[1] The factor by which the CN rms noise was multiplied 
to obtain the cutoff used to make the CO moment maps. For example, for
AM 2055-425, the 
CN moment 0 map was made with a cutoff of $2 \times 1.9 = 3.8$ mJy
beam$^{-1}$ while the CO moment 0 map was made with a cutoff of
$15 \times 2 \times 1.9 = 57$ mJy beam$^{-1}$. See \S\ref{data} for
more details. 
\\
\end{table*}

\subsection{The ALMA Cycle 0 CN galaxy sample}

For this archival ALMA project, we focused on Cycle 0 data, which were
the majority of the public ALMA data at the time we began the project. 
Cycle 0 data have the additional advantage that calibrated uv data sets
are available through the ALMA archive, thus removing the need to
reapply the calibration to the raw uv data. This recalibration step
can be time consuming and require powerful computers with plenty of
disk space, particularly for data
taken before the introduction of the ALMA calibration pipeline
or for data that cannot be calibrated using the pipeline.

The first step was to manually inspect the titles and, when necessary,
the abstracts of all one hundred Cycle 0 projects to identify those
targeting nearby ($z < 0.02$) extragalactic sources. The ALMA Science
Archive query tool 
was then used to confirm which of these extragalactic projects
observed the CO $J=1-0$ transition. We identified a total of 6 suitable projects
(Table~\ref{table-projects}): 5 projects observed a single galaxy while
the 6th project observed 20 galaxies and detected CO emission in 14
of them \citep{ueda2014}. Two of these galaxies (NGC 7135, AM 1158-333) had
sufficiently weak CO emission 
\citep[signal-to-noise (S/N) ratio of $\sim 3-4$,][]{ueda2014}
that it was clear that CN would not be 
detected and were dropped from any further analysis. 


Calibrated uv data and initial image cubes  were downloaded
from the ALMA 
Science Archive in March 2014. 
M83 and NGC 253 were observed using 
mosaics, while the other galaxies were observed with  single ALMA
pointings. After making a preliminary image of the 
CN line in VV 114 to get an idea of the typical CO/CN ratio, we then 
inspected the CO and CN data cubes for each of
the 17 galaxies to measure the S/N ratio at the 
peak of the CO emission and to look for CN emission with S/N $>3$. 
8 galaxies showed significant CN emission across more than one
velocity channel. These galaxies comprise the ALMA Cycle 0 CN galaxy sample and 
are listed in Table~\ref{table-sample}.
The remaining galaxies were not detected in CN or, in the case of NGC
7252,  have only a marginal CN
detection in a single velocity channel and are discussed in
Appendix~\ref{app-upperlimits}.  

\subsection{Initial data processing}

Continuum subtraction was performed for all the galaxies in the ALMA
Cycle 0 CN sample except  NGC 1377.
Cleaned data cubes of the CO $J=1-0$ and the CN 
113.4910 GHz emission were made for all 8
galaxies. 
The clean box
was defined by inspecting emission channel by channel for each line.
For VV 114, 
which was imaged in both the compact and extended
configurations, we use only the compact data as we judged that their
higher surface brightness sensivitity would make it easier to detect the
weak CN lines. 
CO and CN data cubes for NGC 253 were kindly
provided by A. Bolatto, while a cleaned CO data cube for VV 114 was
kindly provided by K. Sliwa. 

For each galaxy, the velocity ranges with strong CO emission 
were inspected to search for $> 3 \sigma$ emission in the CN line.
The presence of hyperfine structure around the CN line makes it
difficult to measure the CO/CN ratio as a function of velocity without
additional spectral processing. For example,
the very high signal-to-noise ratio in NGC 253 means the CN line
actually appears more 
extended in velocity space than the CO emission. This is due to the
21 MHz ($\sim 55$ km s$^{-1}$) spread in the CN hyperfine structure
around the strongest component of the line.
Therefore, although the spectral cubes
were used for initial inspection of the data, the primary analysis of
the CO/CN line ratio was done using integrated intensity maps.
Before making the integrated intensity (moment 0) maps, a small
spatial smoothing was applied to each of the CO and CN data cubes to 
match the beam size and shape (Table~\ref{table-sample}).

\subsection{Measuring the CO/CN ratio}
\label{ratios}

In all the galaxies studied here, the CO emission is much stronger
than the CN emission. As a  result, the CO data cubes can
pick up much more extended emission, both spatially and spectrally,
than the CN data cubes can. This mis-match in sensitivity poses a
challenge for determining the CO/CN ratio accurately. For example, in
AM 2055-425, the CO/CN ratio in individual channels in the cube can be
as low as 10, while the global ratio measured from integrated
intensity maps approaches 50. Although there is clear evidence that the
CO/CN ratio can vary from place to place within a galaxy
\citep{meier2015}, it seems likely that significant variations in this
line ratio can also be caused by the different signal-to-noise (S/N)
ratios in the two lines. 

Consider the simple case of a compact (point-like) galaxy with a true
CO/CN ratio of 10. CN
emission would then be detected with a S/N $\ge 2$ only in those
velocity channels where CO had a S/N $\ge 20$. However, CO would be
detected with a S/N $\ge 2$ in many more velocity channels. This would
result in the measured CN luminosity being an underestimate of the
true CN luminosity, which would result in a measured CO/CN ratio
larger than the true value of 10. However, if we measured the CO
luminosity only from those channels where the CO S/N was $\ge 20$,
then we would recover the true CO/CN ratio of 10.
Therefore, we adopt the following method to measure the 
CO/CN line ratio.

\begin{figure*}
	\includegraphics[width=\columnwidth]{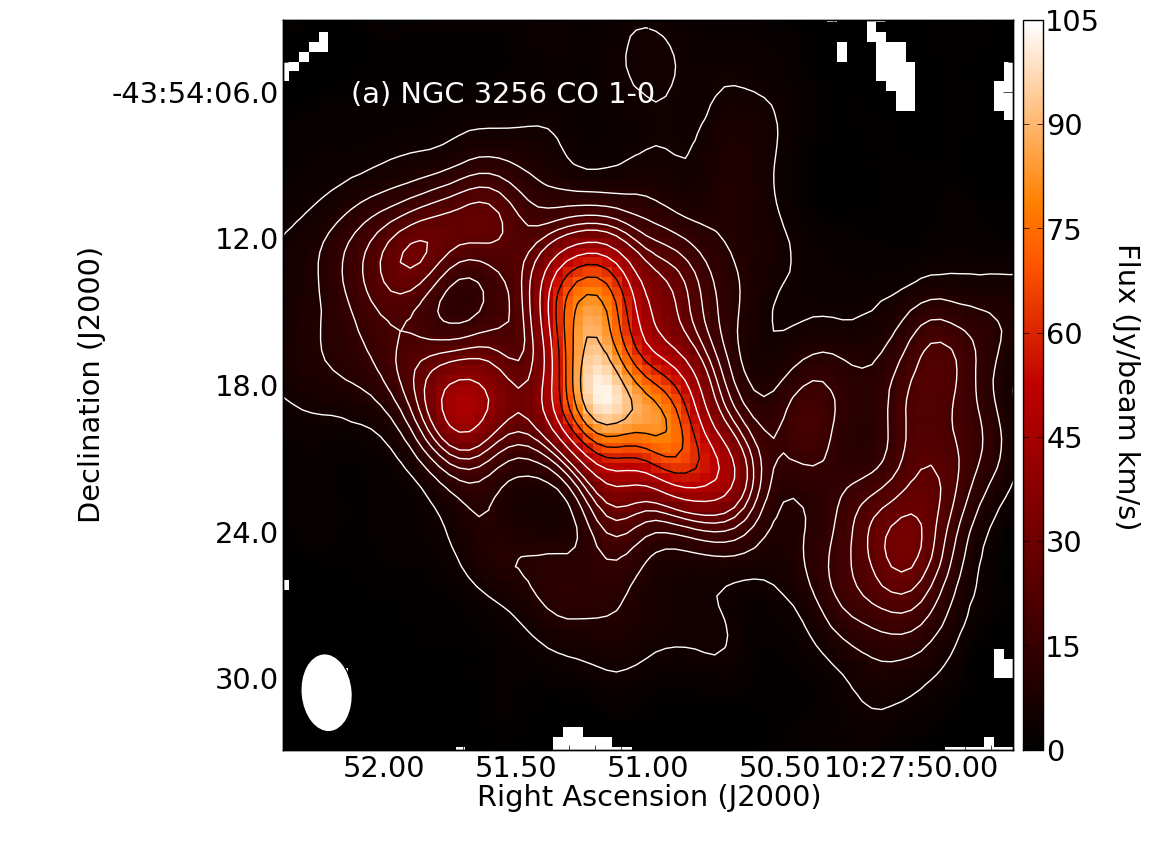}
	\includegraphics[width=\columnwidth]{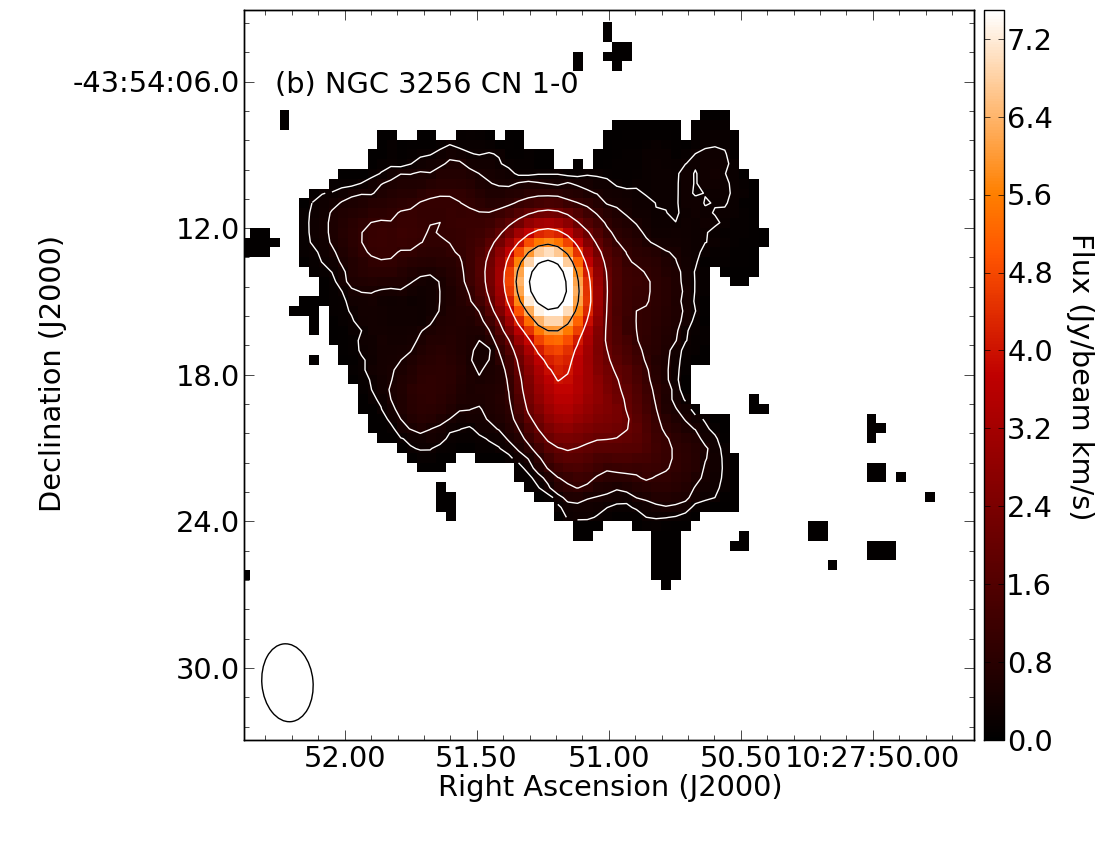}
	\includegraphics[width=\columnwidth]{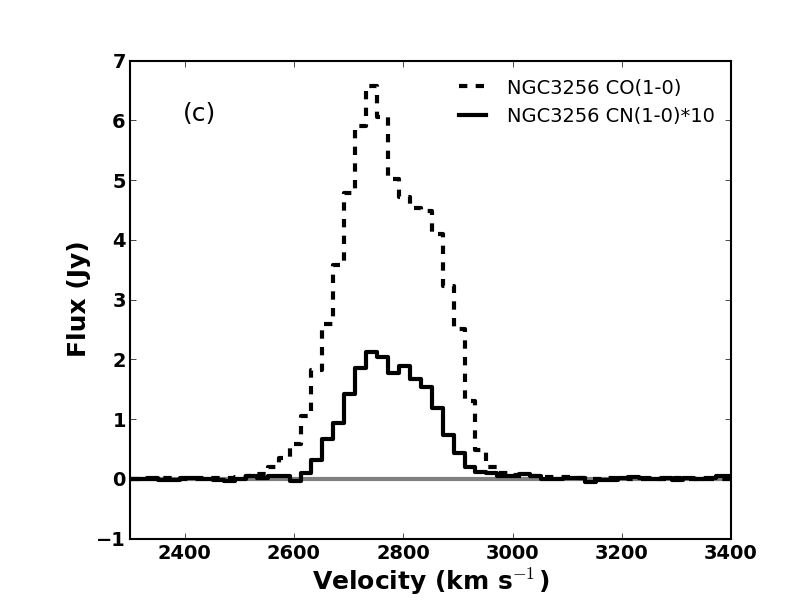}
	\includegraphics[width=\columnwidth]{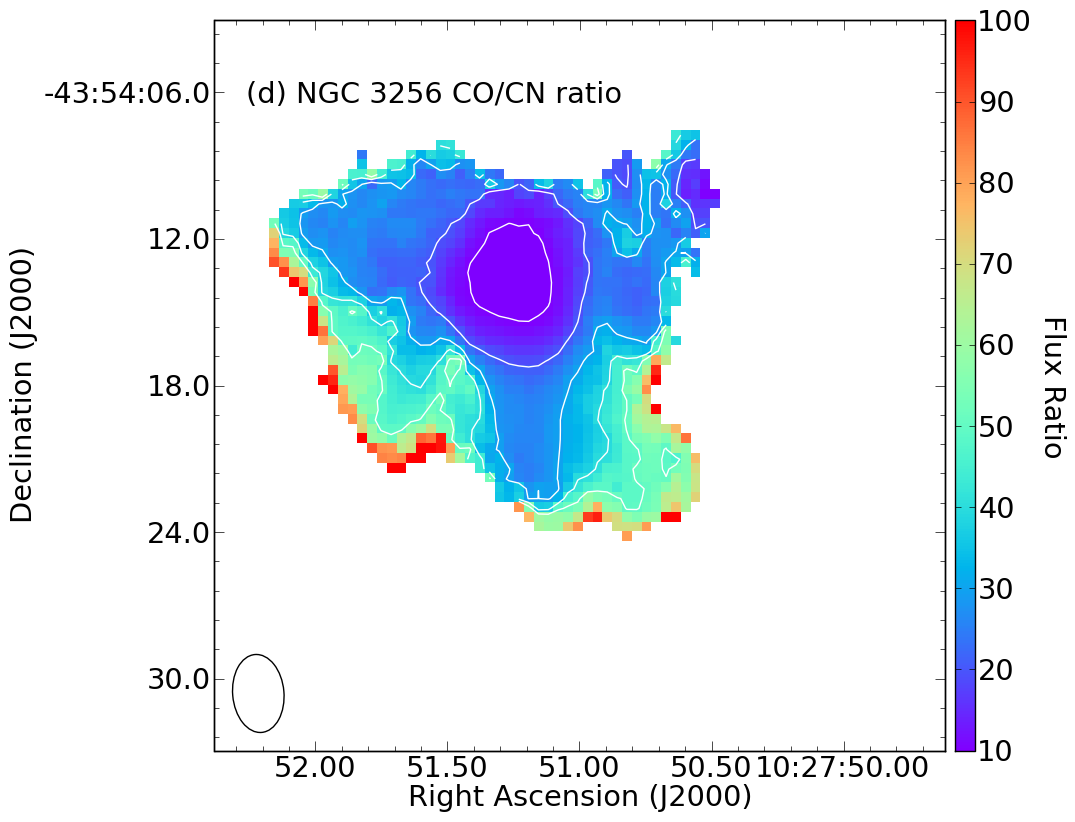}
    \caption{Example of CO and CN processing in NGC 3256. (a) CO
    1-0 integrated intensity image. Contours are 5, 10, 15, 20, 25,
    30, 40 ... 90 Jy
      beam$^{-1}$ km s$^{-1}$. The beam is shown by the ellipse in the
      lower left corner. (b) CN 113.491 GHz integrated
    intensity. Contours are 0.25, 0.5, 1, 2, 4, 6, 8 Jy
      beam$^{-1}$ km s$^{-1}$.  
(c) CO and CN spectra
      integrated over the entire emission region for each line. The CN
    spectrum has been multiplied by a factor of 10. 
(d) CO/CN line ratio with CO masked at 10 times
    CN level. Contours are 10, 20, 30, 40, 50. 
(A ratio map made with CO clipped at 30 times the CN
    level shows only small differences from this map in the outer
    regions.) While the mean CO/CN ratio averaged over this map is 22
    (Table~\ref{table-ratios}), NGC 3256 shows clear spatial variations
    in the CO/CN line ratio, with a minimum value towards the 
    CN peak near the northern nucleus and a value 3
    times larger towards the CO peak near the southern nucleus.}
    \label{fig-ngc3256}
\end{figure*}

An integrated
intensity map was made for each line using a $2-3\sigma$ cutoff and
including only those velocity channels in which sufficiently strong emission
was seen (Fig.~\ref{fig-ngc3256}). An initial image of the ratio of the two
integrated intensity  maps 
was made using a mask made from a clipped
CN map. For the galaxies with the strongest emission, 
the CN map was clipped at a level equivalent
to 3$\sigma$ emission times two 
velocity channels, while a clip of 2$\sigma$ times two velocity
channel was used for the weaker galaxies (Table~\ref{table-sample}).

The procedure used to obtain a signal-to-noise matched (S/N-matched)
measure of the CO/CN 
line ratio involved iterating on the signal-to-noise cutoff used in
making the CO integrated intensity map. 
An initial mean CO/CN ratio, $R_1$, was estimated
from the first ratio map.
To match the signal-to-noise ratio in the two lines, a second CO
integrated intensity map was made, but this time using a cutoff equal
to $R_1 \times (2,3)\sigma_{CN}$ and using the same velocity channels used
to make the CN integrated intensity map. A new ratio map was made
as before using the second CO moment map, and the mean CO/CN ratio was
remeasured, giving a new ratio $R_2$. This process was repeated 2-3 times
until the value of $R_i$ used in making the CO moment map was
reasonably close to the value of $R_{i+1}$ measured from the resulting
moment map ratio, at which point the process was deemed to have
converged (Fig.~\ref{fig-ngc3256}). 
Finally, CO and CN fluxes were measured from the two final moment 0 maps
after applying the same mask used in the ratio map. The S/N-matched
CO/CN line ratio is given by the ratio of these two fluxes. 

We illustrate this technique for one of the galaxies in our sample, AM
2246-490. The initial CO/CN ratio map showed values around 10 in the
centre of the map. A new CO moment 0 map was made with a flux cutoff
of $10\times 
2\sigma$; however, the new CO/CN ratio map showed much
lower values with a mean of around 2. A third CO moment map was made
using a flux cutoff of $5\times 2\sigma$, which gave CO/CN ratio map
with a mean of around 7. A final CO moment map was made
using a flux cutoff of $6\times 2\sigma$, which also gave CO/CN ratio map
with a mean of around 7. The flux-weighted CO/CN ratio for this galaxy
was then measured to be $7\pm1$ (see Table~\ref{table-ratios}). We
note that the agreement between the flux-weighted CO/CN ratio and the
value measured from the ratio of the moment 0 maps is not always
as good as for AM 2246-490.

\begin{figure*}
	\includegraphics[width=\columnwidth]{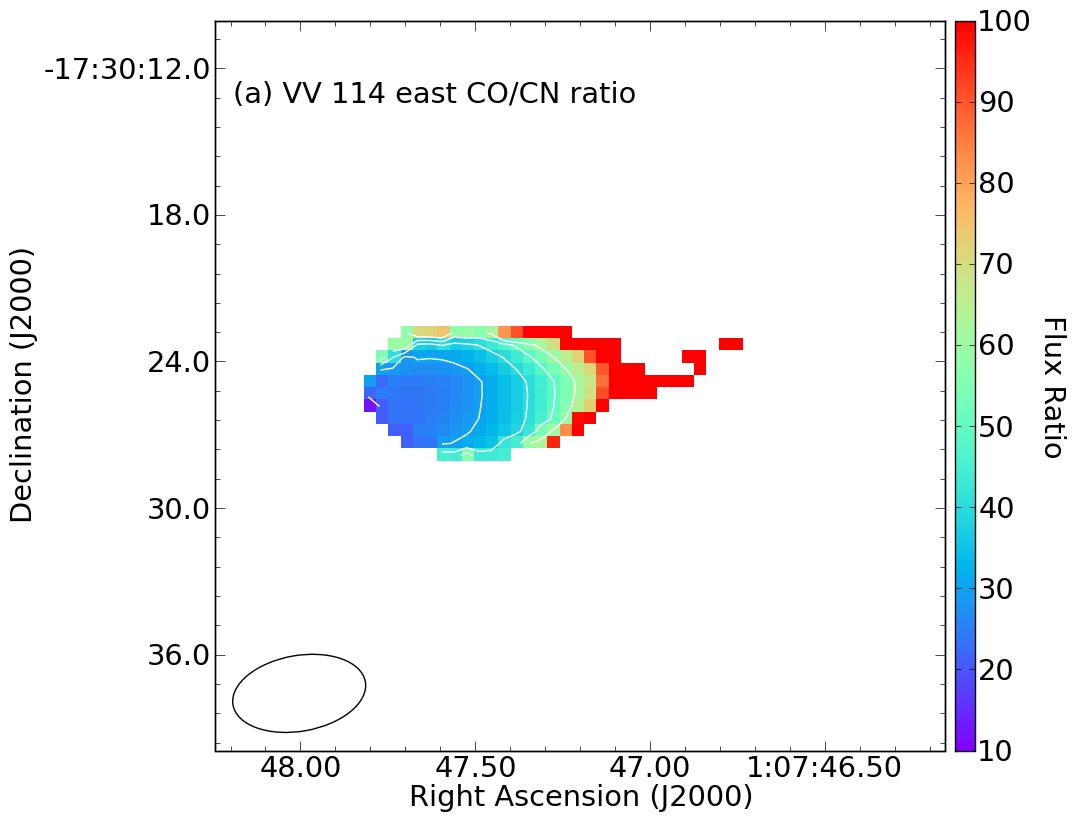}
	\includegraphics[width=\columnwidth]{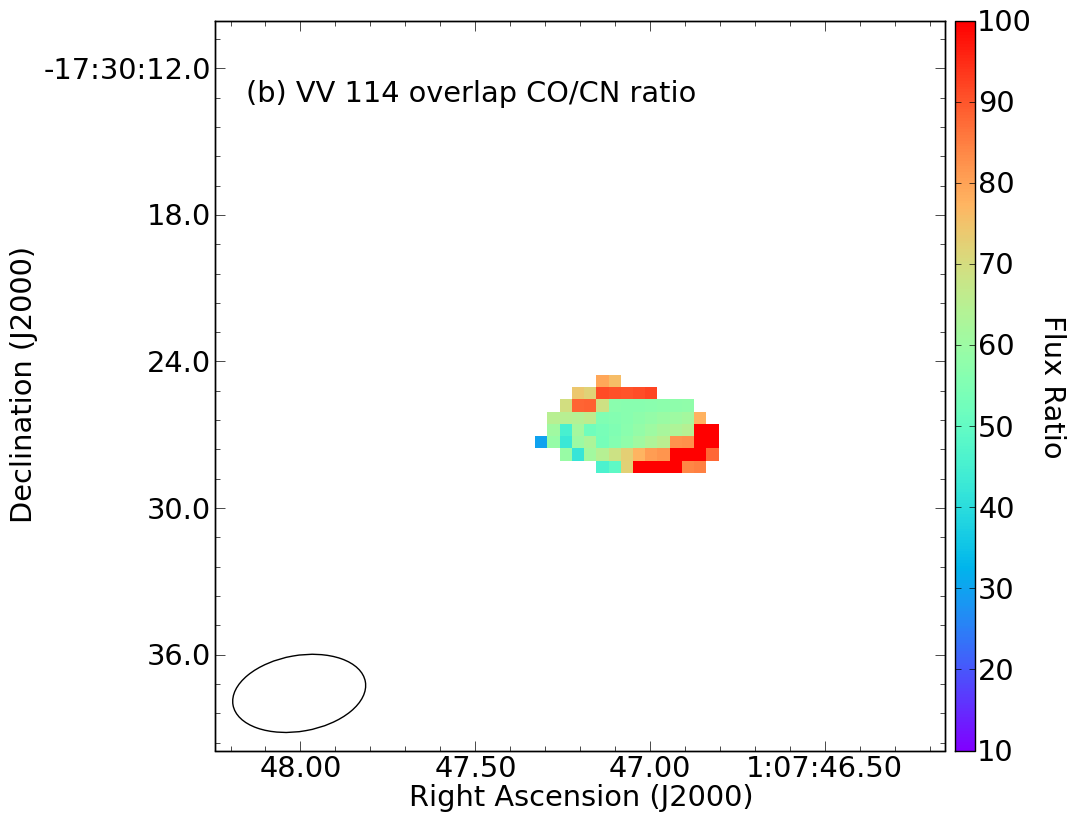}
    \caption{The CO/CN line ratio in VV 114 shows 
      large spatial variations. (a) The CO/CN ratio in the
      eastern nucleus has a mean value of $\sim 30$. Contours are 10,
      20, 30 ... 60. (b) 
    The CO/CN ratio in the overlap region to the west is more than twice as
    large as that in the eastern nucleus. The two images were made
    selecting appropriate velocity 
    channels for each region and clipping the CO moment 0 map to the
    level appropriate for each region; see
    Appendix~\ref{individual-galaxies} for more details.} 
    \label{fig-vv114}
\end{figure*}

The S/N-matched  CO/CN
ratios as well as the 
global ratios measured from integrating the emission over the two
initial integrated intensity maps are given in
Table~\ref{table-ratios}. 
Some notes on the individual galaxies are given in
Appendix~\ref{individual-galaxies}. VV114 is a particularly tricky
case given the large variation in the CO/CN line ratio across the
source. Images of the line ratio tuned for each of the two bright
regions are shown in Fig.~\ref{fig-vv114}.

\begin{table*}
	\centering
	\caption{CO/CN line ratios in nearby galaxies}
	\label{table-ratios}
	\begin{tabular}{llllllllll} 
		\hline
		Galaxy & CO/CN S/N- & CO/CN 
                & $S_{CO}$\footnotemark[1]\footnotemark[2] & $S_{CN}$\footnotemark[1]\footnotemark[2] & $D_L$\footnotemark[3] & 
                $\log (L_{IR})$\footnotemark[4] & Notes\\
& matched ratio\footnotemark[1] & global ratio\footnotemark[1] & (Jy km s$^{-1}$) & (Jy km s$^{-1}$) & (Mpc) & ($L_\odot$) & \\
		\hline
                AM 2055-425 & $14\pm 3$ & $50\pm 10$ & 49 &
                $1.0\pm 0.2$ & 186 & 12.06 & ULIRG, AGN; Fig.~\ref{fig-am2055} \\ 
AM 2246-490 & $7 \pm 1$ & $20 \pm 3$ & 34 & $1.6 \pm 0.2$ & 186 &
11.86 & LIRG; Fig.~\ref{fig-am2246} \\
		NGC 3256 & 22 & 44 & 1646 & 36 & 44 & 11.81 & LIRG
                with an AGN; 
                Fig.~\ref{fig-ngc3256} \\ 
		VV 114 & 65 & 147 & 613 & 4.0 & 82 & 11.69 & LIRG with
                an AGN; 
                Figs.~\ref{fig-vv114},~\ref{fig-vv114images}\\ 
                AM 1300-233 & $13 \pm 2$ & $50\pm 7$ & 89 & $1.7\pm
                0.2$ & 98 & 11.51 & LIRG; Fig.~\ref{fig-am1300} \\ 
                NGC 253 &  16 & 20 & 8575 & 421 & 3.5 & 10.55 &
                starburst; Figs.~\ref{fig-ngc253},~\ref{fig-vv114images}\\ 
                M83  &  $21\pm 1$ & $43\pm 2$ & 979 & $22 \pm 1$ & 4.7 & 10.33 &
                starburst; Figs.~\ref{fig-ngc253},~\ref{fig-m83}\\ 
                NGC 1377  & $40\pm 10$ & $140\pm 40$ & $45\pm 5$  &
                $0.32 \pm 0.08$ & 24 & 10.13& lenticular, AGN,
                 FIR excess; Fig.~\ref{fig-ngc1377}\\ 
		\hline
	\end{tabular}
\\
\footnotemark[1] Calculated uncertainties are given for individual
galaxies where 
the measurement uncertainty on the weak CN emission is the dominant
uncertainty. For galaxies with strong emission, the uncertainty
is due to the exact placement of the measurement apertures and is
 estimated to be at least 5\%.\\
\footnotemark[2] Global CO fluxes measured from ALMA data except for
NGC 253 and M83, for which the ALMA maps cover only a portion of the
galaxy. For both galaxies, CO and CN fluxes were measured in a 30$^{\prime\prime}$ diametre aperture
centred on the central starburst. \\
\footnotemark[3]NGC 253 distance from \citet{radburn-smith2011}; M83
distance averaged from \citet{radburn-smith2011} and \citet{saha2006};  
remaining
galaxies from redshift and NED adopting WMAP 5-year cosmology 
($H_o = 70.5$ km s$^{-1}$ Mpc$^{-1}$, $\Omega=1$, $\Omega_m = 0.27$).\\
\footnotemark[4]Infrared luminosities from \citet{sanders2003}
adjusted to luminosity distance given here. 
\\
\end{table*}


\begin{figure*}
	\includegraphics[width=\columnwidth]{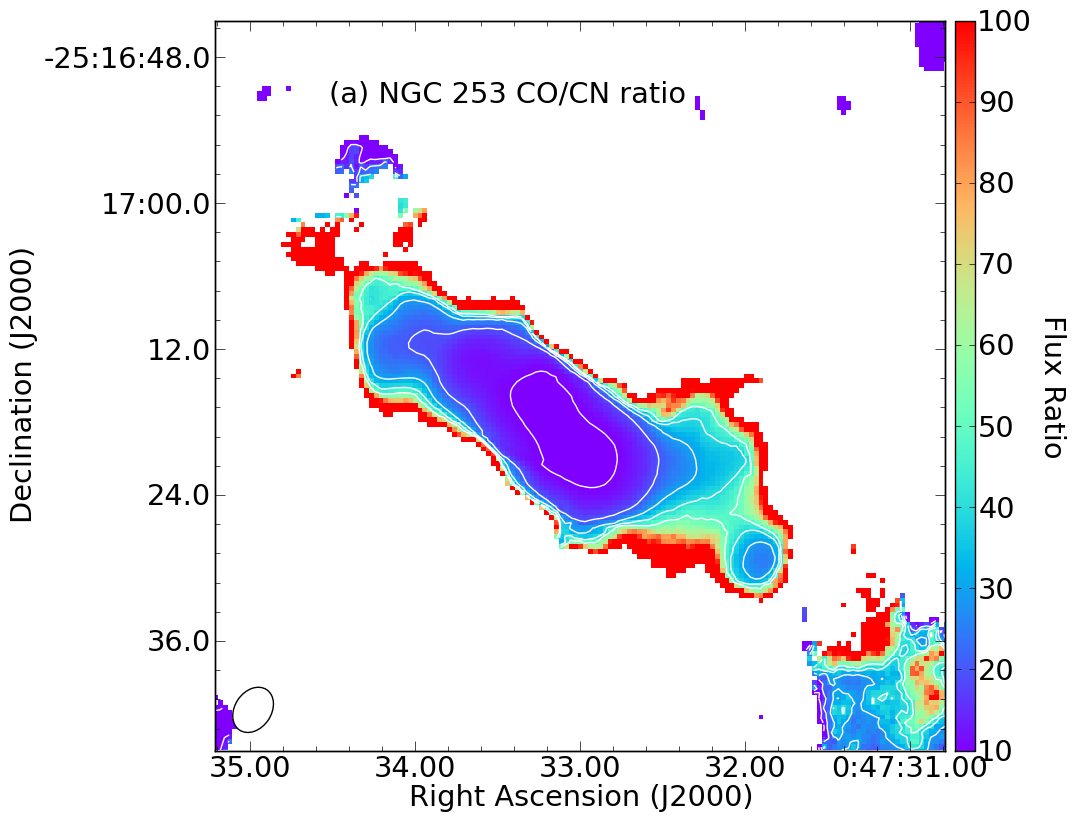}
	\includegraphics[width=\columnwidth]{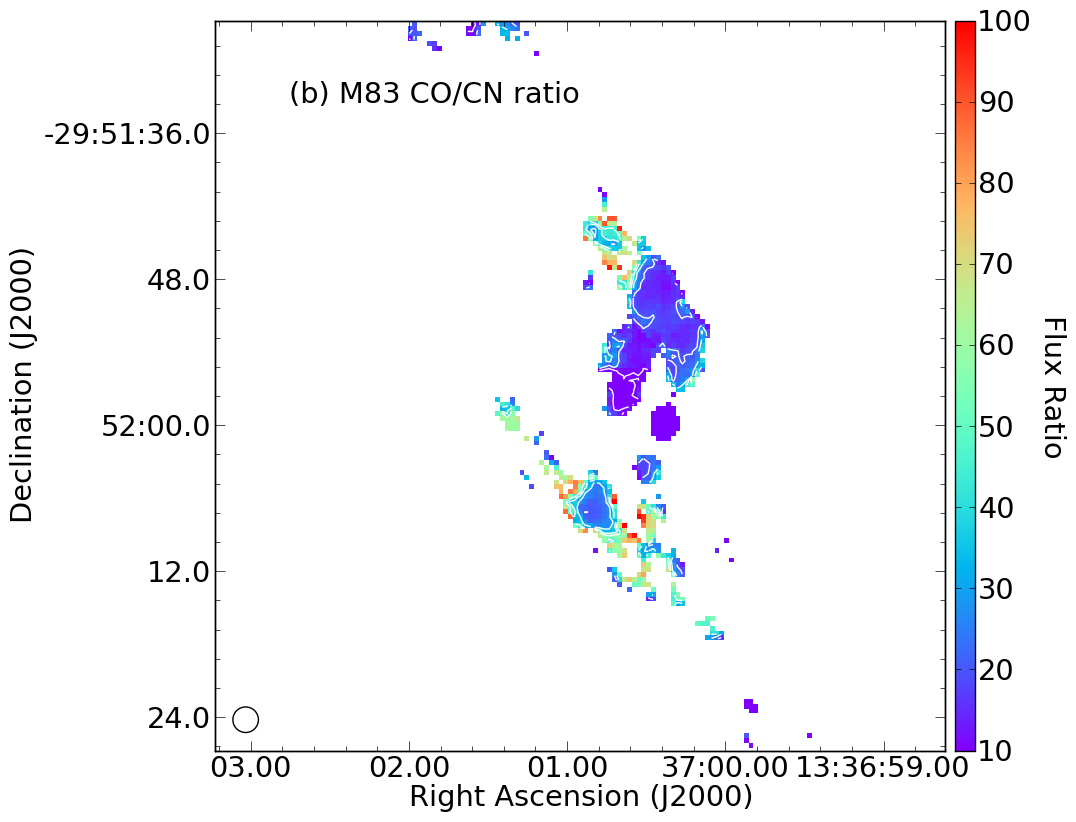}
    \caption{(a) The CO/CN line ratio in the central $1^\prime$ of NGC 253 shows a
      clear radial trend, with values as small as 8 towards the
      central starburst rising to 30-40 in the outer parts
      of the central disk. (b) The CO/CN line
     ratio in the central $1^\prime$ 
      of M83. 
The high resolution of the ALMA data
      (52 pc) resolves individual molecular clouds in this nearby
      spiral galaxy.
 There are clear variations in the CO/CN line
      ratio from cloud to cloud, with the lowest ratios seen in the
      central starburst.}
    \label{fig-ngc253}
\end{figure*}


\section{Large variations in CO/CN within and between galaxies}
\label{discussion}

\subsection{Observed variations in the CO/CN line ratio}

In total, CN emission is detected in 8 galaxies: 5 U/LIRGs, 2 nearby starburst
spirals,  and 1 S0 galaxy. There is clearly a
wide range of morphologies and line ratios in this sample
(Table~\ref{table-ratios}), even for
galaxies that might otherwise be expected to be quite similar.
Three of the galaxies (VV 114, NGC 3256, AM 1300-233) are mergers
or merger remnants with 
$L_{IR}$ and distances that are similar to within  a factor of $\sim
2$. The two more 
luminous galaxies, NGC 3256 (Fig.~\ref{fig-ngc3256}) and VV 114
(Fig.~\ref{fig-vv114},~\ref{fig-vv114images}) show spatially extended emission with 
varying CO/CN line ratios, while 
AM 1300-233 (Fig.~\ref{fig-am1300}) shows compact emission in both lines. 
Interestingly, the
CO/CN line ratios are a factor of two smaller overall in NGC 3256 than
in VV 114.
NGC 3256 is closer than VV 114 and  has been observed with a factor of 3
higher spatial resolution
(Table~\ref{table-sample}). However, averaging over a
similar spatial extent as in the VV 114 maps gives CO/CN line ratios
in NGC 3256 of 14 in the northern 
minimum, which includes the northern nucleus \citep{neff2003}, and 36
in the south-west bright peak, near the southern nucleus. 
Thus, even at similar spatial resolution, NGC 3256 has CO/CN line
ratios that are a factor of 2 lower than those in VV114.

AM 2055-425, AM 2246-490, and NGC 1377 show 
very compact CN emission similar to that of 
AM 1300-233 (Figs.~\ref{fig-am2055}, ~\ref{fig-am2246},
and~\ref{fig-ngc1377}). Two of these 
galaxies have  S/N-matched CO/CN line ratios of 13-14,
while NGC 1377 has a significantly larger line ratio of 40 and AM
2246-490 has a line ratio of 7, the smallest line ratio in the entire sample. 
The two very nearby galaxies, NGC 253
and M83, both show spatially extended CO and CN emission
(Figs.~\ref{fig-ngc253}, ~\ref{fig-ngc253images},
and~\ref{fig-m83}). The emission  
in NGC 253 nearly fills the field of view, while the emission in M83
breaks up into distinct clouds. 
Although some of this difference may be
due to their different inclinations (edge-on versus close to
face-on), the majority is likely due to the factor of 5 higher CO
$J=1-0$ intensity in the centre of NGC 253 compared to M83 \citep{young1995}.

The spatial resolution of these data sets ranges from 50 pc in M83 to
2.1 kpc in VV 114. There are
significant spatial variations in the CO/CN line ratio in NGC 253 and
M83, the two
closest galaxies. Fig.~\ref{fig-ngc253} shows that CO/CN line ratios
as small as 10 
are observed in the central starburst in each galaxy, with ratios
rising to 20-30 over a larger area. To attempt to match the
measurements for the more distant galaxies, the CO/CN line ratios
reported in Table~\ref{table-ratios} are measured over a
30$^{\prime\prime}$ (0.5-0.7 kpc) diametre aperture. For NGC 1377, we
made test images applying a strong taper to the uv-data, which gave a
beam of $\sim 0.5$ pc, similar to that of NGC 3256. The global and
S/N-matched CO/CN line ratios measured from this lower
resolution map agreed extremely well with the values in
Table~\ref{table-ratios}. Thus, we feel that the higher spatial
resolution obtained for M83, NGC 253, and NGC 1377 is not biasing
the measured CO/CN line ratios in a significant way.


\subsection{Why does CN vary? Abundance versus excitation}
\label{merger-variations}

Even in the galaxies with the highest signal-to-noise measurements of
the CN line, the CO/CN line ratio
varies considerably, both from place to place inside a single galaxy
and from one galaxy to another. Spatially resolved line ratios range
from  $\le 10$ in NGC 3256 and NGC 253 to $\ge 70$
in VV114. The primary question is whether variations in the line ratio are due
to changes in the abundance or changes in excitation, and also which
molecule (CN or CO) is primarily responsible for the observed
changes. 
An additional complication is that changes in the optical depth
of CN or CO (due to differences in the H$_2$ column density in
different regions and galaxies) could produce changes in the
observed line ratio as well. We address each of these possibilities
below, focusing primarily on the three brightest galaxies, NGC 3256,
VV 114, and NGC 253.


\subsubsection{Optical depth and the CO-to-H$_2$ conversion factor}


Previous analyses of the CN emission for NGC 3256
\citep{sakamoto2014}, VV 114 \citep{saito2015}, and NGC 253
\citep{meier2015} all showed that the CN $N=1-0$ lines are optically
thin. 
In NGC 253, the ratio of strong/weak CN line ranges from values of  $\sim 2$
consistent with 
optically thin emission in LTE up to values of 3-5 that are consistent
with LTE at the 2$\sigma$ level calculated from \citet{meier2015}.
In VV 114, the CN line emission in the eastern nucleus is also
consistent with optically thin emission \citep{saito2015}.
Thus, it seems unlikely that variations in the CN optical depth can be
responsible for the spatial variations  seen in the CO/CN line ratios.

In principle, variations in the optical depth of the CO $J=1-0$ line
could produce changes in the CO/CN line ratio.
The CO $J=1-0$ line is generally thought to be optically thick.
For NGC 253, \citet{meier2015} concluded from the ratio of the
CO/C$^{17}$O line ratio that the CO $J=1-0$ line is 
moderately optically thick ($\tau \sim 5$). In VV114,
\citet{sliwa2013} find an optical depth of $\sim 14$ in the CO $J=1-0$
line by fitting multiple lines of CO and $^{13}$CO.
For NGC 3256, \citet{sakamoto2014} conclude that the lowest 3
transtions of CO are likely to be optically thick and thermally excited.


Under these conditions, the CO $J=1-0$ integrated intensity is often
used as a direct tracer of the H$_2$ column
density via the CO-to-H$_2$ conversion factor
\citep[e.g.][]{bolatto2013b}.
If this conversion factor changes from galaxy to galaxy, or from place
to place within a galaxy, it could play a role in the observed changes
in the CO/CN line ratio. However, these three strong emission line
galaxies are either starbursts (NGC 253) or LIRGS (NGC 3256, VV114)
and so the lower CO-to-H$_2$ conversion factor seen in
galaxies with high star formation rates 
\citep{downes1998} seems likely to be appropriate for
these galaxies. 

For the specific case of NGC 253, we can examine what effect would be
produced by
a changing CO-to-H$_2$ 
conversion factor from the central starburst to the outer parts of the
nuclear disk. The CO/CN map (Fig.~\ref{fig-ngc253}) shows a ratio of
10 in the centre rising to perhaps 40 in the outer disk. If the
CO-to-H$_2$ conversion factor varied from the starburst value (0.8 $M_\odot$
pc$^{-2}$) to the
standard spiral value (3.2 $M_\odot$
pc$^{-2}$) \citep{bolatto2013b}  over the
same region, this variation 
would strengthen the inferred radial variation in the H$_2$/CN ratio,
rather than weakening it. \citet{meier2015} adopt a constant starburst
conversion factor across the entire central region of NGC 253
and we will do the same in this analysis.

Similarly, the differences in the CO/CN line ratios between NGC 3256
and VV114 also seem unlikely to be due to changes in the CO emission.
\citet{rosenberg2015}
investigate the CO spectral line energy distributions (SLEDs) for a
number of LIRGs, including VV 114 and NGC 3256. 
The total
luminosities of the CO J=4-3 to J=12-11 lines in 
the two galaxies agree to within 20\%, suggesting the CO excitation in
the warm molecular gas in the two galaxies is very
similar. 
Table~\ref{table-ratios} shows that 
the CO $J=1-0$ luminosity in VV 114 measured from the ALMA map 
is roughly 30\% larger than that in
NGC 3256, but this difference in CO luminosity is not enough to
account for the difference in the CO/CN line ratio on its own.
Thus, we conclude that the effects of the CO optical depth cannot
explain the changes in the CO/CN line ratio observed in these galaxies.




\subsubsection{Excitation versus abundance for CN}

For the optically thin CN
line, the integrated intensity, $I$, is proportional to
$N_{CN} / (T_{ex} \exp{(E_u/k T_{ex})})$, where $N_{CN}$ is the column
density of CN, $E_{u}$ is the energy of the upper state, and
$T_{ex}$ is the excitation temperature of the line
\citep[cf.][]{meier2015}. The 
relatively low value of $E_u$ for the CN $N=1-0$ line 
(equivalent to  5.45~K) implies that under many
conditions $I \propto N_{CN}/T_{ex}$. However, the large critical
density of this line \citep[$\log n = 6.24$,][]{meier2015} means that
the excitation 
temperature of CN may be lower than the kinetic temperature of the gas.

For this analysis, we assume that the CO line intensity is proportional to the H$_2$
column density. Under this assumption, an increase in the CO/CN line ratio could be due to a
decrease in the CN column density and hence in the average CN abundance
relative to H$_2$, [CN]/[H$_2$]. Alternatively, an increased CO/CN
line ratio could be due to an increase in the excitation
temperature of the CN line. 
The excitation temperature is constrained to be less than or equal to
the kinetic temperature of the gas, and thus a larger excitation
temperature implies an increase in the 
kinetic temperature and/or the density of the molecular gas. 
We first discuss theoretical predictions for the CN abundance in PDRs
and then discuss constraints on the CN excitation. We defer a
discussion of the CN abundance in XDRs to \S~\ref{sec-agns}.


\citet{boger2005} have made detailed PDR models of the CN and HCN
abundance for a variety of gas densities ($n$), ultraviolet radiation
fields ($\chi$), and cosmic ray ionization rates. They use a plane parallel PDR code with
updated rate coefficients for some key reactions and
model the abundances and integrated column densities to a depth of
$A_v = 10$ ($N_H \sim 2\times 10^{21}$ cm$^{-2}$). We discuss their
results for $A_v = 10$ as being most relevant to an ensemble of
molecular clouds in a gas-rich starburst galaxy.
In their models, the CN abundance ($N_{CN}/A_v$) has a minimum at
$n=10^4$ cm$^{-3}$ and increases by a factor of
3 for densities of 10$^6$ cm$^{-3}$ and by a factor of 10 for
densities of 10$^3$ cm$^{-3}$ (with the
ionization parametre $\chi/n=0.2$ held constant). 
These same models show that
the CN/HCN ratio increases by a factor of 30 for densities
increasing from 10$^4$ cm$^{-3}$ to 10$^6$ cm$^{-3}$. Increasing the
ionization parametre by a factor of $10^3$ produces a factor of 2
decrease in the CN abundance and a factor of 4 decrease in the
CN/HCN ratio. Finally,
an increase of a factor of 10 in the cosmic ray ionization rate increases
the CN abundance by a factor of 10, the HCN abundance by a factor of 5
and the CN/HCN ratio by a factor of 3.


\citet{meijerink2007} compare PDR and XDR models for a range of
densities and radiation fields. Their PDR models include recombination
reactions on polycyclic aromatic hydrocarbons (PAHs), which has the
effect that the transition from C$^+$ 
to C occurs at larger column densities \citep{meijerink2005}.
In their PDR models, the CN/HCN column
density ratio
depends primarily on the gas density and decreases by a factor of 3 for
densities increasing from 10$^4$ cm$^{-3}$ to 10$^6$ cm$^{-3}$. This
is in the opposite sense to the trend in
\citet{boger2005}. 
\citet{meijerink2011} explore the effect of
increased cosmic ray heating on the HCN emission for two different values
of the gas density and radiation field strength. At high densities
($10^{5.5}$ cm$^{-3}$),
the models show an 
increase of a factor of 3 in the HCN abundance for an increase of a
factor of $10^3$ in the cosmic ray ionization rate, while at low densities
($10^3$ cm$^{-3}$) a
decrease by a factor of 4 is seen over the same change. These changes
are generally smaller than the effect of cosmic rays seen in
\citet{boger2005}. Unfortunately, the CN abundance is not discussed in
\citet{meijerink2011} and the CN/$A_v$ ratio is not shown in
\citet{meijerink2007}, which limits the usefulness of these models
for our current analysis.

\citet{ginard2015} compare their high resolution observations of CN in
M82 to a grid of models using the Meudon PDR code. They use a
constant density ($4\times 10^5$ cm$^{-3}$) with 2 different values each for the
ultraviolet radiation and cosmic ray ionization rate and 5 different cloud sizes
with different central $A_v$. They find that the CN
abundance increases with increasing cosmic ray ionization rate, although not as
strongly as in \citet{boger2005}. Their models also show  that the CN
abundance increases 
slightly as the strength of the ultraviolet radiation field increases;
in contrast, \citet{boger2005} saw a slight decrease with increasing
radiation field.

These three sets of PDR models show that the abundance of CN can vary
by factors of  a few with changing density, ultraviolet
radiation field, and cosmic ray ionization rate. 
For example, the models of 
\citet{boger2005} imply that the high CN abundance (low
CO/CN ratio) in AM 2246-490 (as compared to AM1300-233 and AM
2055-425) could be produced by an increase in the
cosmic ray ionization rate, a decrease in the ultraviolet radiation field, and/or
either a high ($10^6$ cm$^{-3}$) 
or low ($10^3$ cm$^{-3}$) gas density.
However, the three models do not
always agree on the sense of the change that is expected, which makes
it difficult to draw firm conclusions without observations of
additional CN transitions. 
In addtion, only the models of \citet{boger2005} are able to produce
an order of 
magnitude change in the CN abundance of the kind seen when comparing
AM 2246-490 with the VV114 overlap region. Observations of HCN, for
which the predicted trends also differ between the models, could
be particularly 
helpful in constraining the parametres of the gas.

The CO/CN line ratios for the resolved and/or nearby galaxies in our sample
cluster around values of $\sim 10$ in starburst regions (the centres
of M83 and NGC 253 and the northern nucleus of NGC 3256).
Somewhat larger values of 20-30 are seen in the
extended disk emission of these 3 galaxies. The bright
CO emission combined with a very high CO/CN ratio of $\sim 60$ in the 
overlap region in VV114 stands out as particularly unusual. (Although
ratios this large are seen in the outskirts of NGC 3256 and to some
extent in NGC 253, the CO intensity is much weaker in those regions.) This ratio
is 4-6 times
larger than the ratios measured in the starburst nuclei in our sample.
If such a large CO/CN line ratio is due to a low CN abundance, it
could be produced by a low cosmic ray ionization rate
\citep{boger2005,meijerink2007,ginard2015}, a high ultraviolet
radiation field \citep{boger2005}, a gas density of around $10^4$
cm$^{-1}$ \citep{boger2005}, or smaller clouds
\citep{ginard2015}.

An alternative explanation for the large CO/CN
line ratio seen in VV114 is a high CN excitation temperature. 
High excitation temperatures typically
require a high kinetic temperature, which can be produced
by strong UV radiation fields
\citep{meijerink2011}, possibly combined with a high density.
These conditions would imply
a high local gas pressure, which 
suggests that measurements of the gas pressure might be a useful tool
to distinguish between the effects of CN abundance and CN excitation.
The CN $N=1-0$ line has a critical density that is fairly similar to
that of the CO J=6-5 line. Although this CN line is only 5 K above
ground (compared to 116 K for CO J=6-5), the astrochemical models suggest
that CN will be found with high abundance in warm UV or X-ray irradiated gas.
This comparison suggests that it is
reasonable to consider constraints for the warm
molecular gas derived from high J CO lines in interpreting the observed CN
emission. 

\citet{kamenetzky2017} have modeled the global CO
excitation from $J=1-0$ to J=13-12 for a sample of 87 galaxies. 
All of
the galaxies discussed here except AM 2238-490 are included in their
sample. The 
pressures of the cold gas for these 7 galaxies agree within their
uncertainties, while 
there are significant variations in the pressure of the warm
component, with NGC 1377 having a warm gas pressure that
is nearly 2 orders of magnitude larger than the mean value.
However, for VV114 and NGC 3256, they
find very similar pressures in  the warm molecular gas, although the
uncertainty on the pressure  in VV 114 is
quite large. 
However, we need to be careful not to
over-interpret these global similarities between the galaxies given
the spatial variations in the CO/CN ratio 
seen in Figs~\ref{fig-ngc3256} and~\ref{fig-vv114}. 
Spatially resolved
observations of higher J rotational transitions of 
molecules with ALMA to measure the warm gas pressure as well
as observations of higher frequency lines of CN would be very helpful
in distinguishing between these different scenarios.

\subsection{The role of  AGN}\label{sec-agns}


An AGN will produce an X-ray dominated region which will alter the
astrochemical balance compared to what is found in photon dominated
regions \citep{meijerink2007}. 
\citet{imanishi2009} show that the HCN/HCO$^+$ ratio can be enhanced
in galaxies that show mid-infrared signatures of an AGN. 
The XDR models of \citet{meijerink2007} show a strong decrease in
CN/HCN with increasing 
density and much larger CN/HCN ratios overall compared to the PDR
models. Four of the galaxies in our sample show evidence for an AGN
and elevated CO/CN line ratios of around $\sim 30$ are seen in 3
of them: the southern nucleus of
NGC 3256, the eastern nucleus of VV114, and NGC 1377.

NGC 1377 has been shown to contain a buried AGN 
\citep{costagliola2016} which is generating a jet as well as 
a slower molecular outflow \citep{aalto2016,aalto2017}. 
It has a CO/HCN ratio of 4.4 and an
HCN/HCO$^+$ ratio $> 1.8$ 
\citep{imanishi2009}, which suggests that the HCN emission is
enhanced relative to both CO and HCO$^+$. NGC 1377 has one
of the largest S/N-matched CO/CN ratios in our sample
(Table~\ref{table-ratios}), 
suggesting a depletion of CN emission near the AGN that could be related
to the enhancement of HCN. 

The eastern nucleus of VV114 has been suggested to contain an AGN by
\citet{iono2013} and also shows the highest
HCN/HCO$^+$ line ratio in the entire system
\citep{imanishi2007}.
The eastern nucleus has a similar CO/CN ratio (Fig.~\ref{fig-vv114})
to the ratio measured in NGC 1377.
Note that the western nucleus of VV114 has
only relatively weak CO emission \citep{saito2015} and is not detected
in CN.

\citet{sakamoto2014} identified molecular outflows from both of the
nuclei in NGC 3256. They argue that the southern nucleus hosts a
(possibly dormant) AGN based on differences in the outflow properties
between the two nuclei. \citet{ohyama2015} used infrared and X-ray
data to identify a heavily absorbed, low luminosity AGN in the
southern nucleus of NGC 3256. The area around the southern nucleus of
NGC 3256 (Fig.~\ref{fig-ngc3256}) has a higher CO/CN line ratio 
($\sim 26$, Appendix~\ref{individual-galaxies}) than the northern nucleus,
consistent with the results for NGC 1377 and VV 114 east.
In contrast to NGC 1377 and VV114, \citet{harada2018} find a {\it lower}
HCN/HCO$^+$ line ratio as 
well as a lower HCN/$^{13}$CO abundance ratio in
the southern (AGN) nucleus than in the northern (starburst) nucleus.

AM 2055-425 has been classified as an AGN/starburst
composite \citep{imanishi2010}. \citet{imanishi2017} find that
infrared radiative pumping plays a role 
in the excitation of HCN in AM 2055-425 and observe high-velocity
wings in the CO J=3-2 emission indicative of a strong molecular outflow.
It has a  S/N-matched CO/CN
line ratio that is significantly lower than those of NGC 1377 or VV114 east.
As the
most distant galaxy in our sample, it is possible that we do not have
sufficient spatial resolution or sensitivity to see clear evidence of
an AGN signature in the CO/CN line ratio. 

This analysis suggests that higher CO/CN line ratios tend to be found
in the vicinity of
an AGN. 
To match these observations, 
the AGN would need 
to reduce the overall CN emission while simultaneously enhancing the HCN
emission. However, the analysis by \citet{meijerink2007} shows a 
significant {\it increase} in the CN/HCN column density for an XDR compared
to a PDR at a given gas density. Those same models also produce
slightly larger HCN/CO $J=1-0$ line ratios in PDRs than in XDRs for the
same gas density.
Recent models by \citet{harada2013} of abundances in an AGN
nuclear disk show that CN/CO can be enhanced in the XDR layer while
HCN/CO is enhanced in the warm midplane and the CN/HCN ratio is
enhanced in the colder outer parts of the disk. However, the resulting
column densities are highly dependent on the details of the disk
models, such as the thickness of the disk \citep{harada2013}.
Expanding
the sample of galaxies with resolved CO, CN, and HCN maps would allow
 better statistics on the effect of an AGN on the CO/CN
line ratio and would provide an interesting comparison with XDR models.

\section{Conclusions}
\label{conclusions}

This paper presents the results of an archival project using ALMA 
Cycle 0 data to survey 
the CO/CN ratio in 17 nearby galaxies. We identified a sample of 8 galaxies
with distances $< 200$ Mpc for which both the CN $N=1-0$ and CO $J=1-0$ lines
were detected. These galaxies range from nearby
starbursts (M83, NGC 253) to ultra/luminous infrared galaxies 
(AM 1300-233, AM 2246-490, AM 2055-425, NGC 3256, VV 114) to an S0
galaxy with an unusual 
far-infrared excess (NGC 1377). 

We measure S/N-matched CO/CN line
ratios as well as ratios calculated from the total flux detected in each
line with ALMA.
The S/N-matched CO/CN line ratios range from as low as 7 to as 
high as 65, while the ratios based on the total detected flux 
range from 20 to 140. In addition, spatial
variations in the CO/CN line ratio of greater than a factor of 3 are seen in 
several of the galaxies with the brightest emission lines.

We discuss several possible explanations for the observed variations
in the CO/CN 
line ratio. We argue that the variations are unlikely to be due to differences
in the opacity of either CO or CN, or to changes in the CO-to-H$_2$ conversion
factor in these infrared-luminous systems. Given the low optical depth of
the CN line, a high CO/CN line ratio can be produced by a high CN excitation
temperature or a low [CN]/[H$_2$] abundance ratio, or both. Additional
resolved measurements to constrain the warm gas pressure and the CN 
line excitation should be able to distinguish between these two possibilities.

Of the 8 resolved nuclear regions in our sample, 4 nuclei (AM 2055-425, 
NGC 1377, NGC 3256 south and VV 114 east) contain evidence for an AGN.
The CO/CN line ratios in 3 of these 4 nuclei appear to be
a factor 
of 2-3 larger than the line ratios seen in starburst-dominated regions;
the fourth galaxy (AM2055-425) may be too distant for us to separate
clearly the AGN and starburst emission.
This result suggests that CN emission is suppressed in the presence of
an AGN, but 
whether this is due to excitation or abundance differences is unclear.
Studying a larger sample of galaxies with and without AGN will show
how significant this trend is; given the large and growing ALMA public
archive, it should be possible to increase the sample of galaxies
with both CO and CN detections by a factor of 4 or more compared to
the sample discussed in this paper.

\section*{Acknowledgements}

We thank the anonymous referee for detailed comments that
significantly improved the content of this paper.
This paper makes use of the following ALMA data:
ADS/JAO.ALMA\#2011.1.00099.S,
ADS/JAO.ALMA\#2011.1.00172.S,
ADS/JAO.ALMA\#2011.1.00467.S,
ADS/JAO.ALMA\#2011.1.00525.S,
ADS/JAO.ALMA\#2011.1.00645.S,
ADS/JAO.ALMA\#2011.1.00772.S.
ALMA is a partnership of ESO
(representing its member states), NSF (USA) and NINS (Japan), together
with NRC (Canada), MOST and ASIAA (Taiwan), and KASI (Republic of
Korea), in cooperation with the Republic of Chile. The Joint ALMA
Observatory is operated by ESO, AUI/NRAO and NAOJ. 
The National Radio
Astronomy Observatory is a facility of the National Science Foundation
operated under cooperative agreement by Associated Universities, Inc. 
This research has made use of the NASA/IPAC Extragalactic Database
(NED) which is operated by the Jet Propulsion Laboratory, California
Institute of Technology, under contract with the National Aeronautics
and Space Administration. 
CDW acknowledges financial support from the 
Canada Council for the Arts through a Killam Research Fellowship. The
research of CDW is supported by grants from the Natural Sciences and
Engineering Research Council of Canada and the Canada Research Chairs
program. CDW would like to thank NRAO for its hospitality and
financial support during the initial and final phases of this project,
Kaz Sliwa and Alberto Bolatto for sharing their data, and Chelsea
Sharon for providing feedback on the revised version of this paper.




\bibliographystyle{mnras}
\bibliography{cycle0CNproject_v3} 





\appendix

\section{Notes on individual galaxies with detected CN emission}
\label{individual-galaxies}

\subsection{AM 1300-233, AM 2055-425, and AM 2246-490}

\begin{figure*}
	\includegraphics[width=\columnwidth]{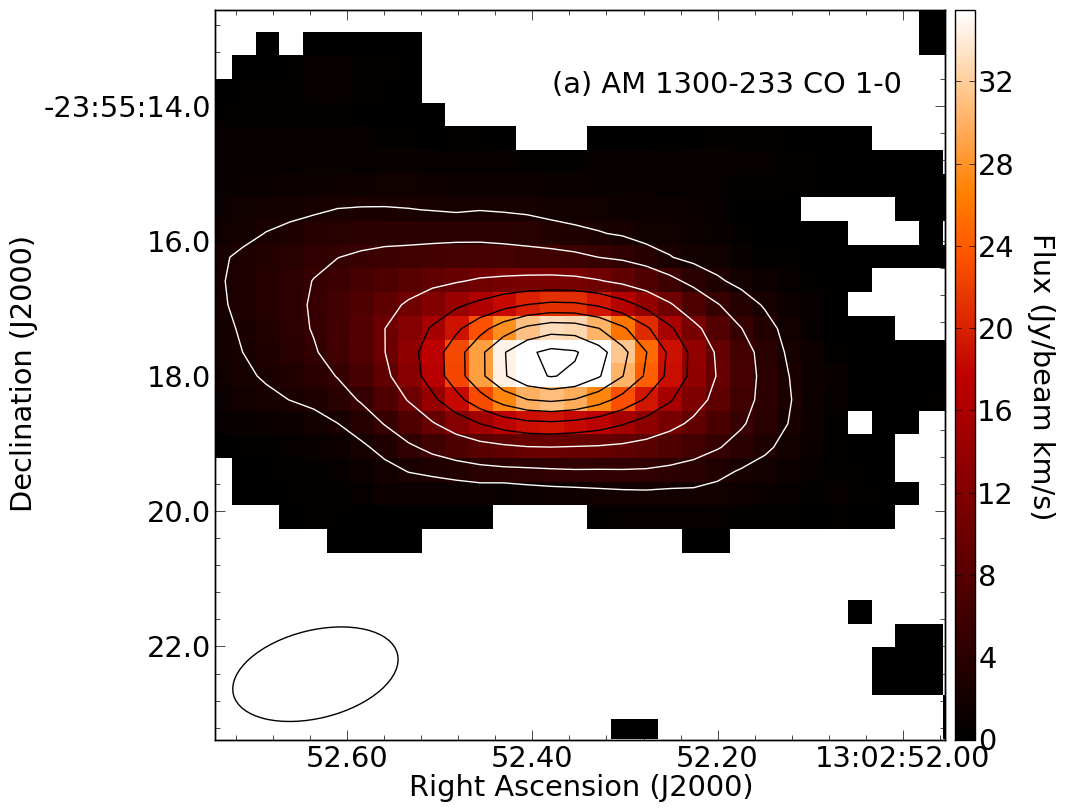}
	\includegraphics[width=\columnwidth]{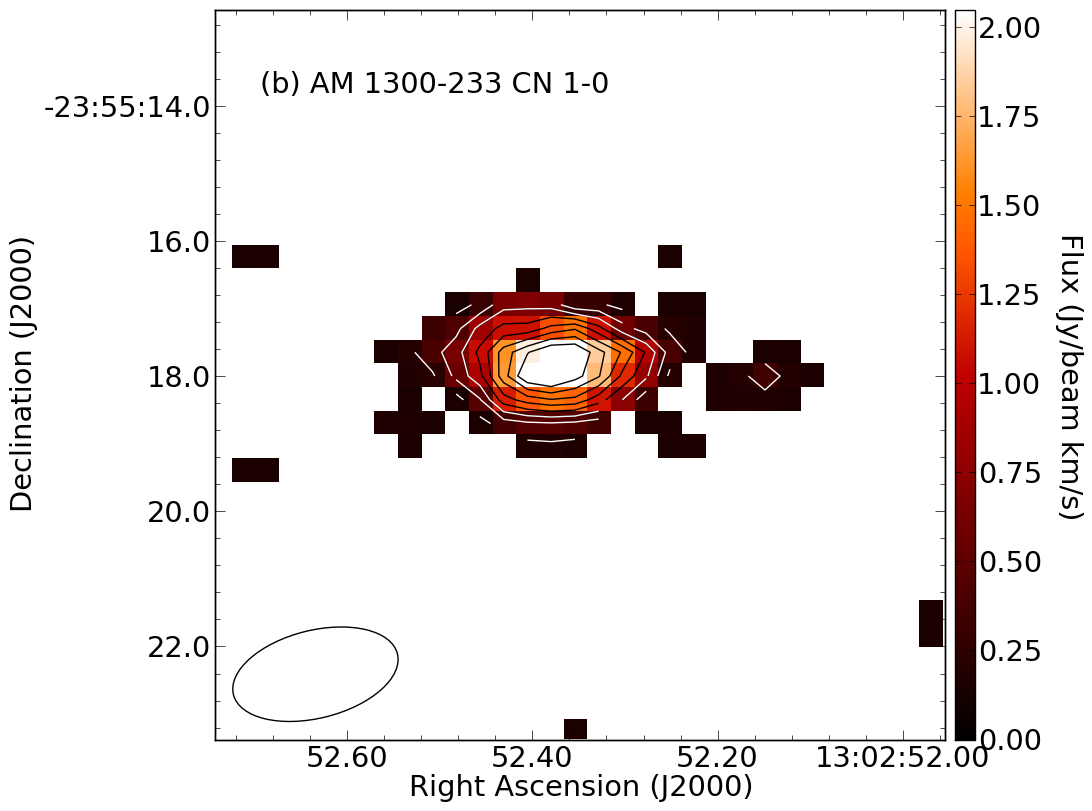}
	\includegraphics[width=\columnwidth]{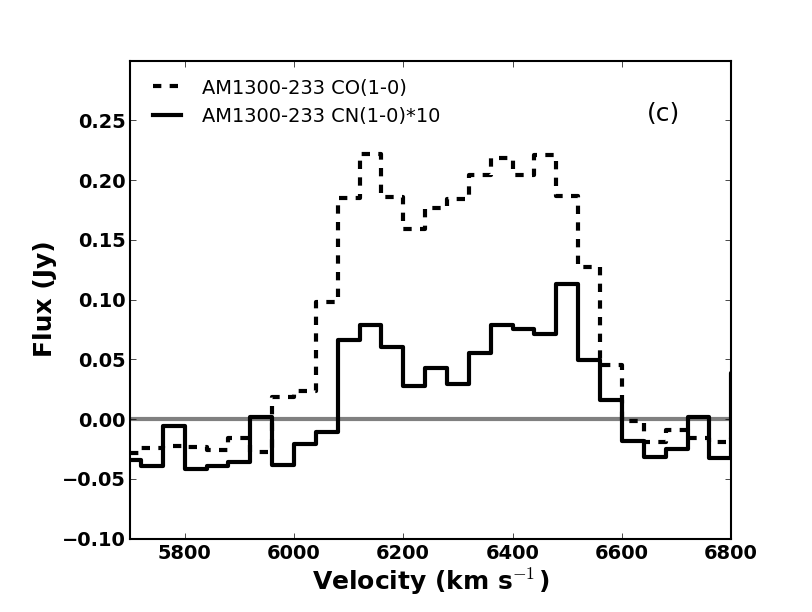}
	\includegraphics[width=\columnwidth]{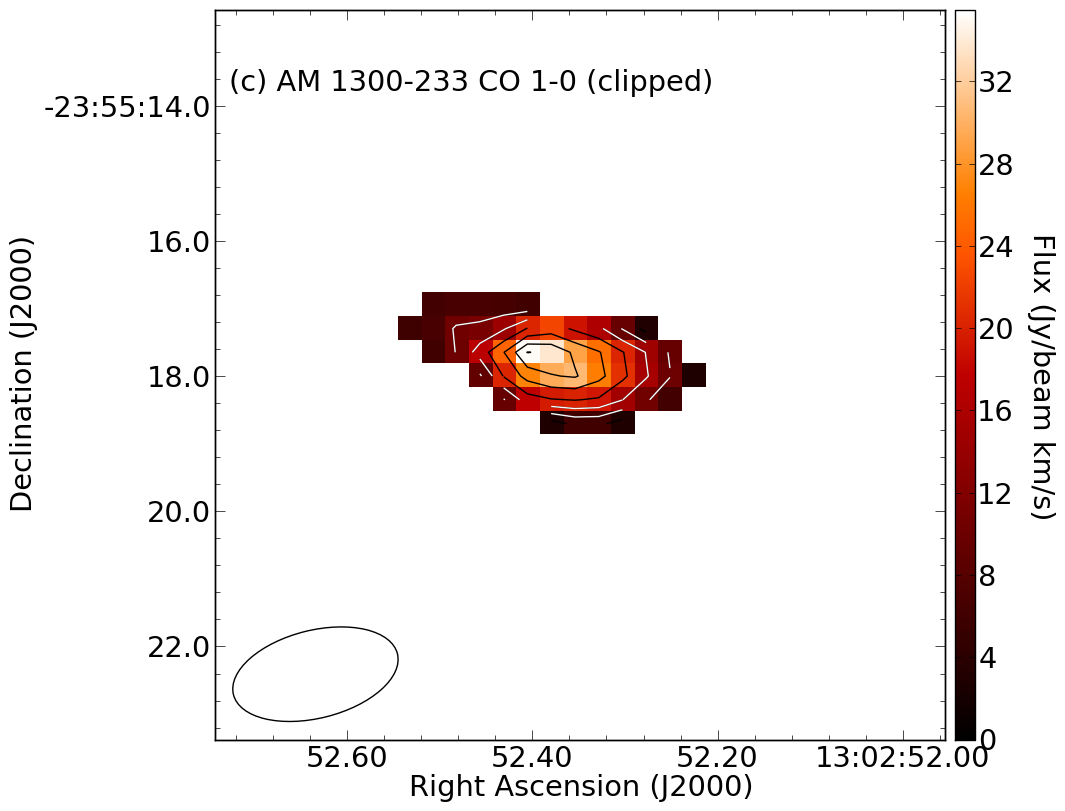}
    \caption{CO $J=1-0$ and CN $N=1-0$ data for AM 1300-233: (a) CO
      integrated intensity image. Contours are 2.5, 5, 10, 15, ... 40 Jy
      beam$^{-1}$ km s$^{-1}$. 
The beam is shown by the ellipse in the
      lower left corner. (b) CN
      integrated intensity image. Contours are 0.25, 0.5, .75, ... 2 Jy
      beam$^{-1}$ km s$^{-1}$. (c) CO and CN spectra
      integrated over the entire emission region for each line. The CN
    spectrum has been multiplied by a factor of 10. (d) CO integrated intensity image
      clipped to match the S/N of the CO image; contours are 10, 15, ... 30 Jy
      beam$^{-1}$ km s$^{-1}$. }
    \label{fig-am1300}
\end{figure*}

\begin{figure*}
	\includegraphics[width=\columnwidth]{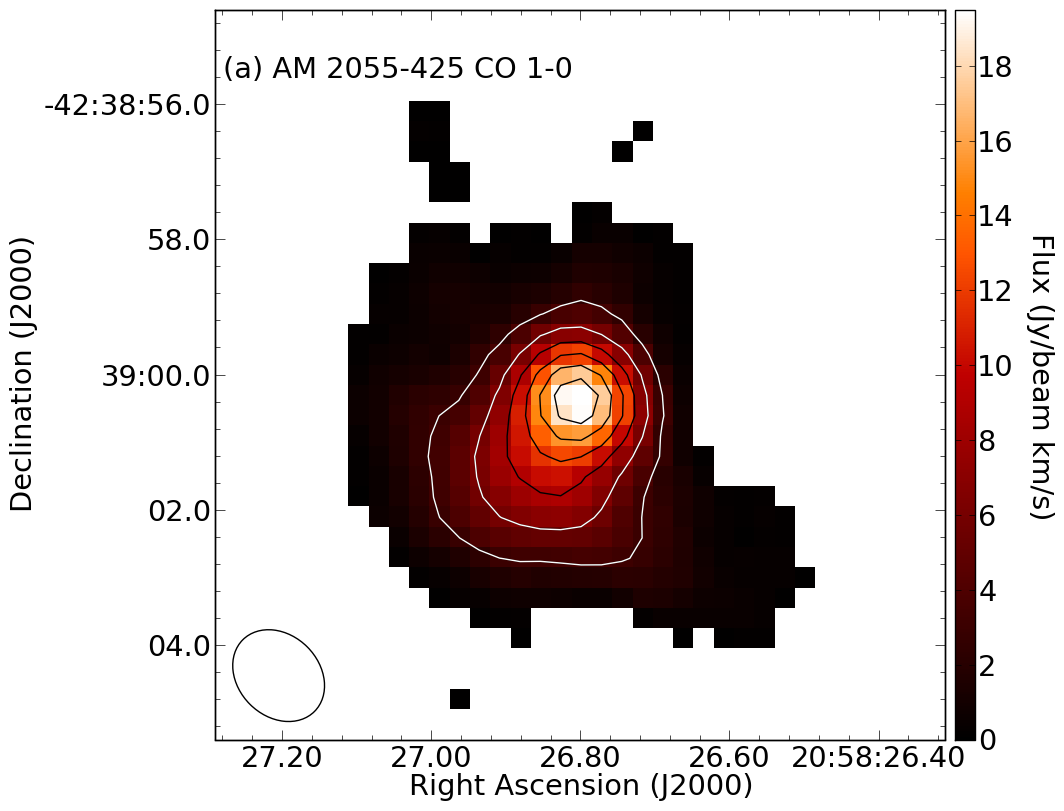}
	\includegraphics[width=\columnwidth]{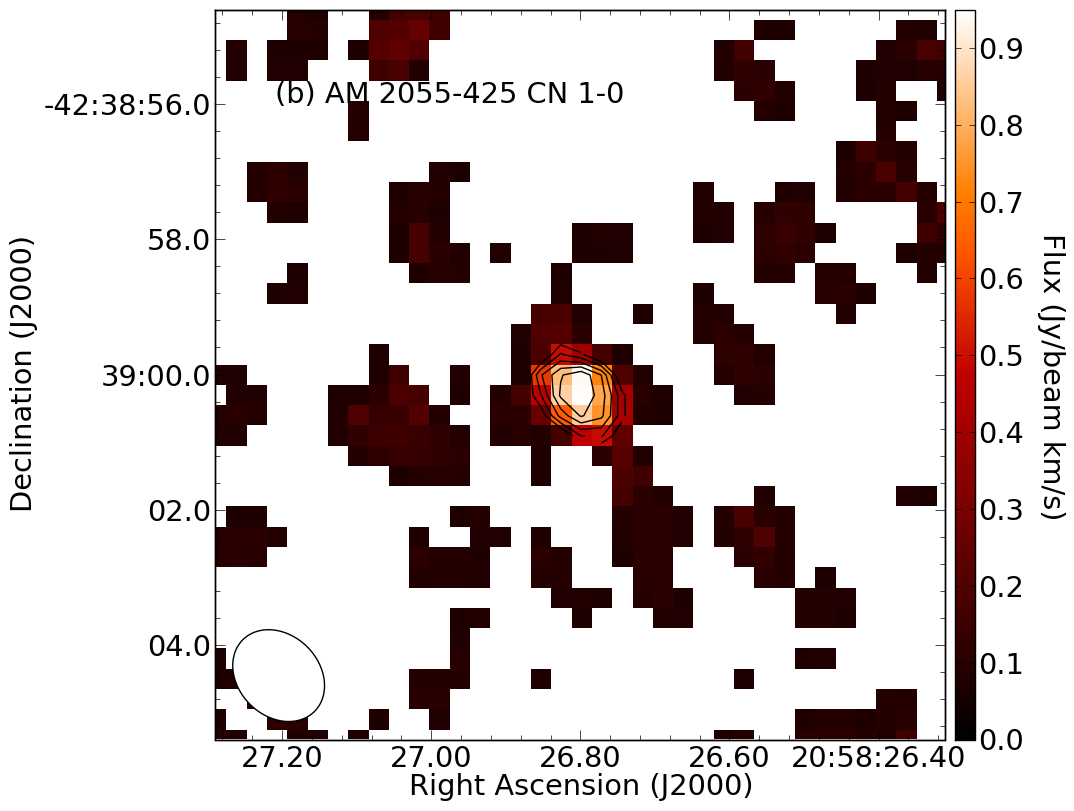}
	\includegraphics[width=\columnwidth]{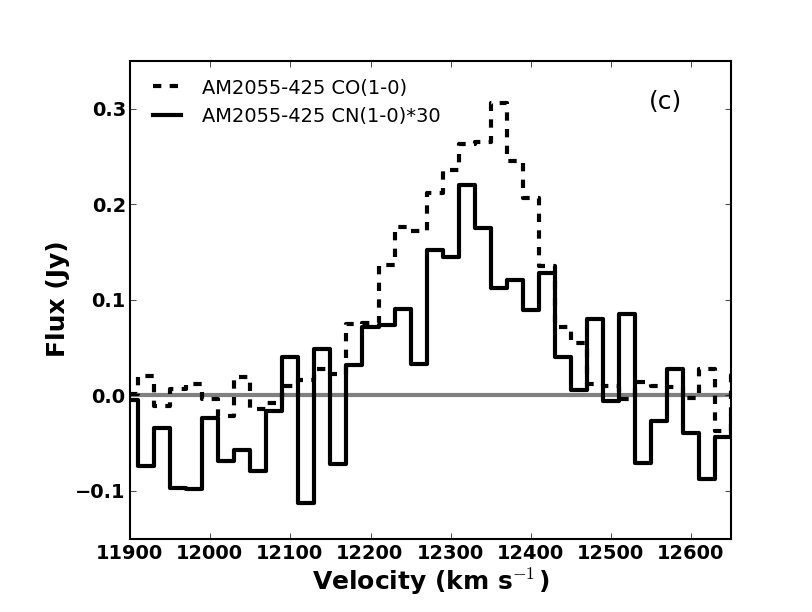}
	\includegraphics[width=\columnwidth]{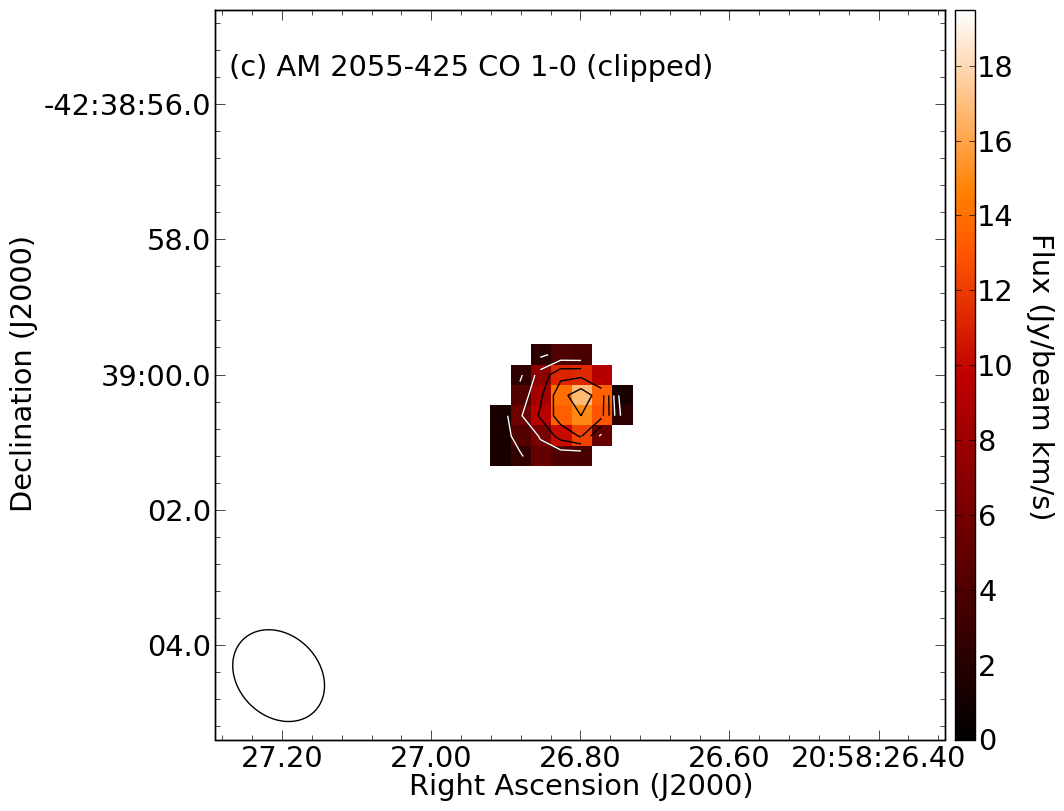}
    \caption{CO $J=1-0$ and CN $N=1-0$ data for AM 2055-425: (a) CO
      integrated intensity image. Contours are 3, 6, 9, ... 18 Jy
      beam$^{-1}$ km s$^{-1}$. The beam is shown by the ellipse in the
      lower left corner. (b) CN
      integrated intensity image. Contours are 0.3,  0.6, 0.9 Jy
      beam$^{-1}$ km s$^{-1}$. (c) CO and CN spectra
      integrated over the entire emission region for each line. The CN
    spectrum has been multiplied by a factor of 30. (d) CO integrated intensity image
      clipped to match the S/N of the CO image; contours are the same
      is in panel (a).  }
    \label{fig-am2055}
\end{figure*}

\begin{figure*}
	\includegraphics[width=\columnwidth]{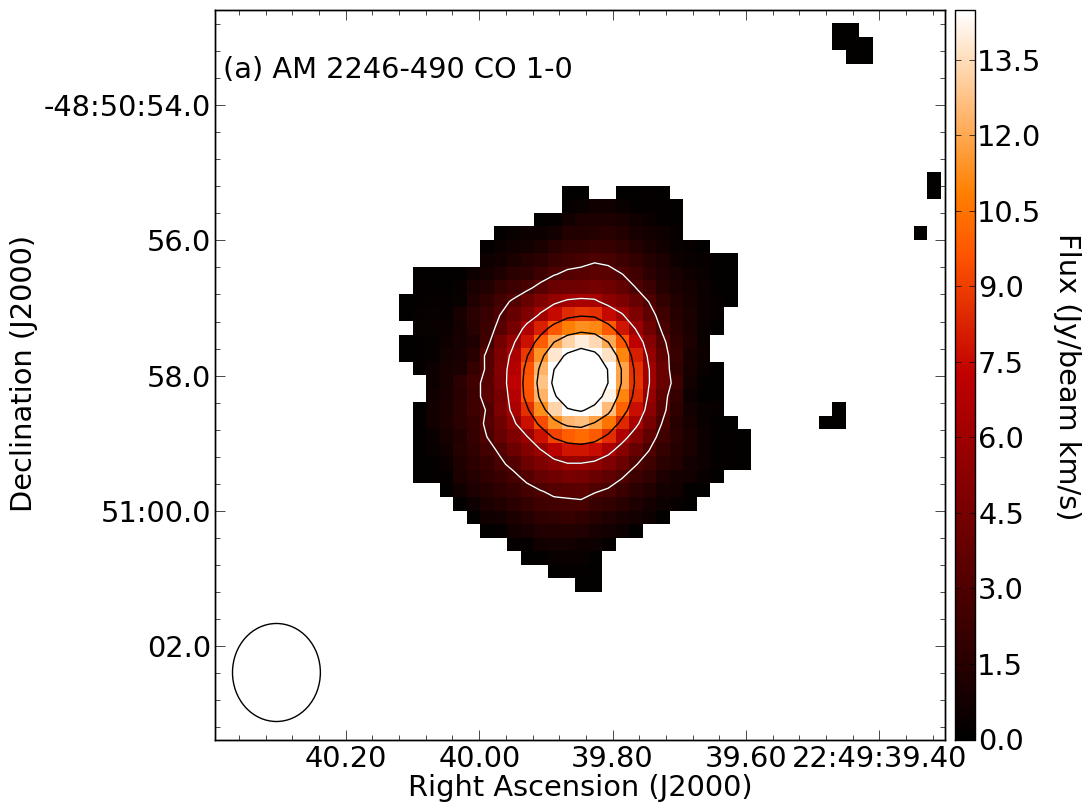}
	\includegraphics[width=\columnwidth]{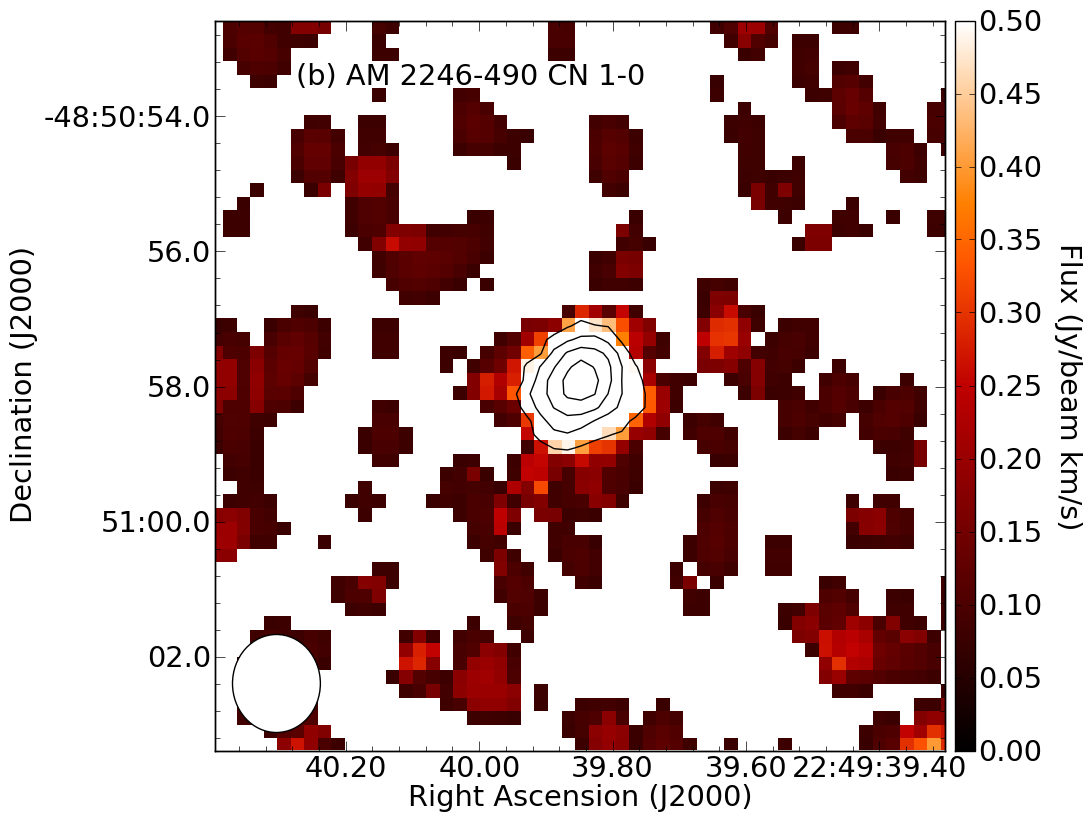}
	\includegraphics[width=\columnwidth]{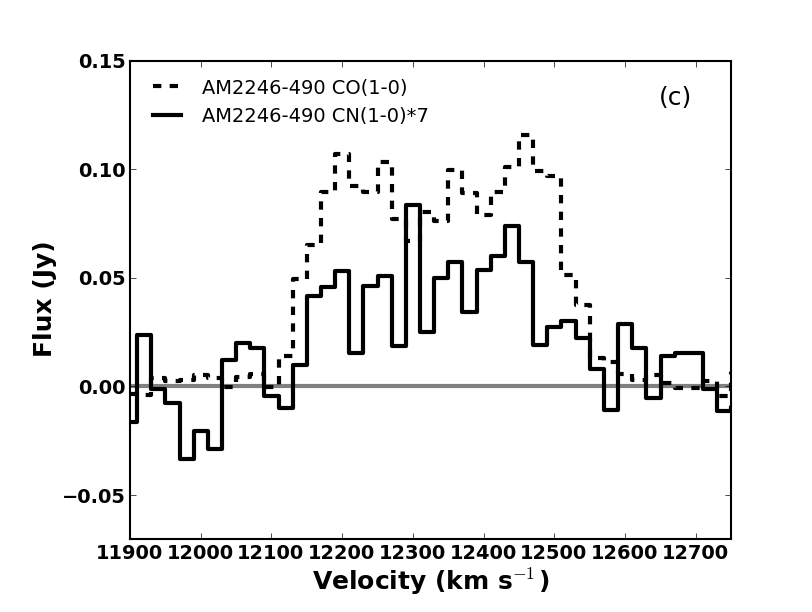}
	\includegraphics[width=\columnwidth]{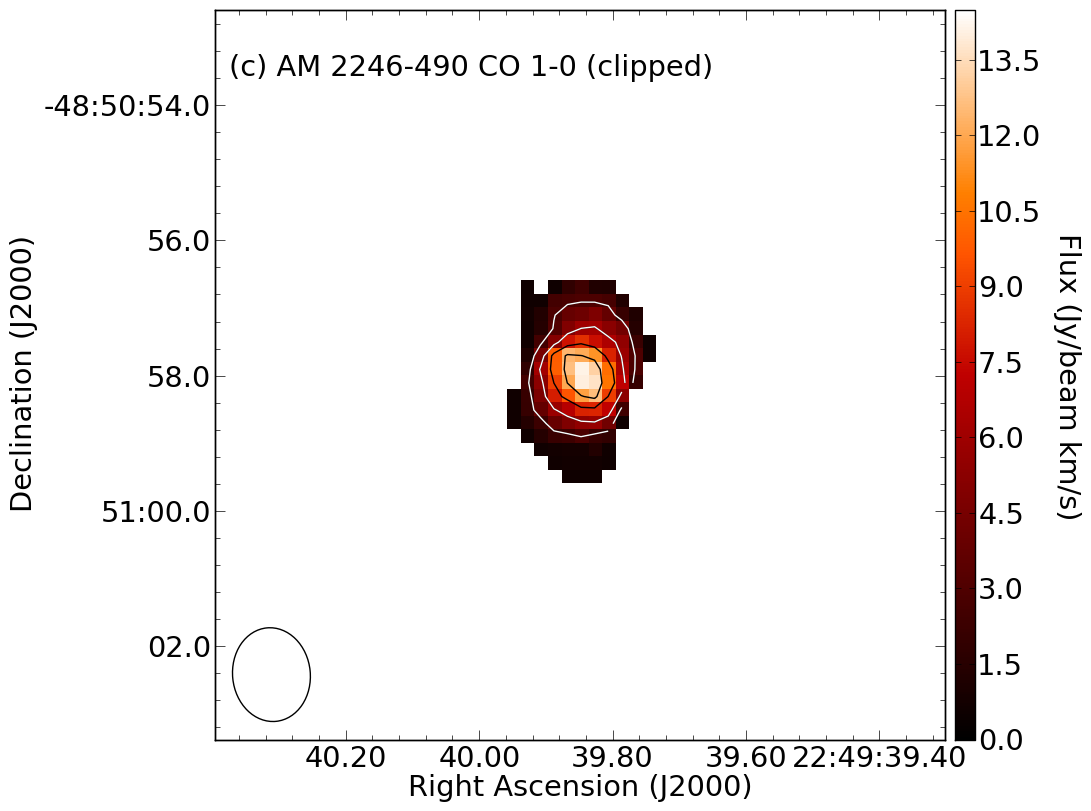}
    \caption{CO $J=1-0$ and CN $N=1-0$ data for AM 2246-490: (a) CO
      integrated intensity image. Contours are 3, 6, 9, 12, 15 Jy
      beam$^{-1}$ km s$^{-1}$. The beam is shown by the ellipse in the
      lower left corner. (b) CN
      integrated intensity image. Contours are 0.45, 0.9, 1.35, 1.8 Jy
      beam$^{-1}$ km s$^{-1}$. (c) CO and CN spectra
      integrated over the entire emission region for each line. The CN
    spectrum has been multiplied by a factor of 7. (d) CO integrated intensity image
      clipped to match the S/N of the CO image; contours are the same
      is in panel (a). }
    \label{fig-am2246}
\end{figure*}

AM 2055-425 has been classified as an AGN/starburst
composite,  while AM 2246-490 and AM 1300-233  are LIRGs with no sign of an AGN
or an obscured AGN in their near-infrared spectra
\citep{imanishi2010,iwasura2011}. 
Although the CN emission for these galaxies
is  quite weak, CN emission does appear at all
the strong CO peaks in the image cubes, and so the 
S/N-matched CO/CN ratio
measurements are likely fairly robust.
However, AM 2055-425 and AM 1300-233 have 
the lowest CO signal-to-noise ratios in the sample (as 
measured in the CO peak emission channel). Thus the global CO/CN ratio
for these galaxies 
is likely to be somewhat overestimated because of the low S/N on the CN
line. Images and global spectra for each galaxy are shown in
Figs.~\ref{fig-am1300},~\ref{fig-am2055}, and~\ref{fig-am2246}.

AM 2246-490 is distinguished by having the smallest CO/CN line ratio
($7\pm 1$) in our entire sample, implying the CN emission is bright
relative to CO.
\citet{stierwalt2014} 
find a relatively low ratio of PAH luminosity to infrared luminosity
($L_{PAH}/L{IR}$), which they say is consistent with its
classification as a merger remnant in
\citet{ueda2014}. \citet{diaz-santos2013} report a low ratio of [CII]
to far-infrared luminosity, which is consistent with its relatively
large infrared luminosity. They also find that AM 2246-490 lies at the
warm colour end of their sample of LIRGs.

\subsection{M83}

\begin{figure*}
	\includegraphics[width=\columnwidth]{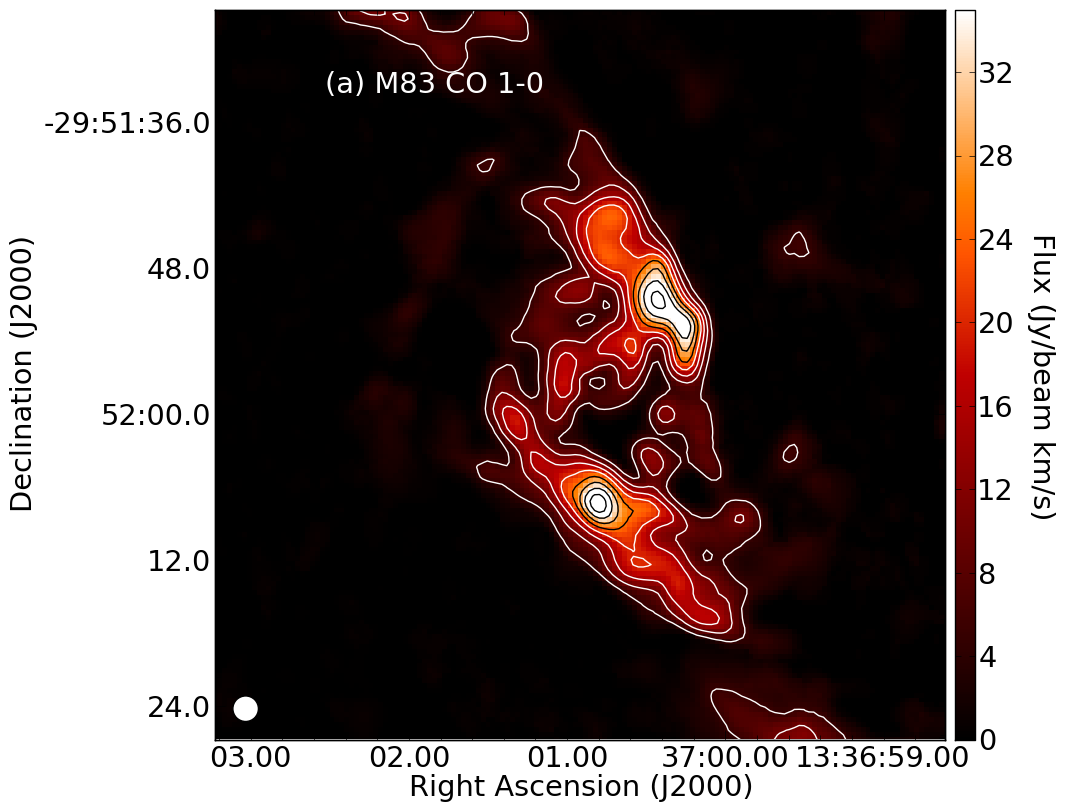}
	\includegraphics[width=\columnwidth]{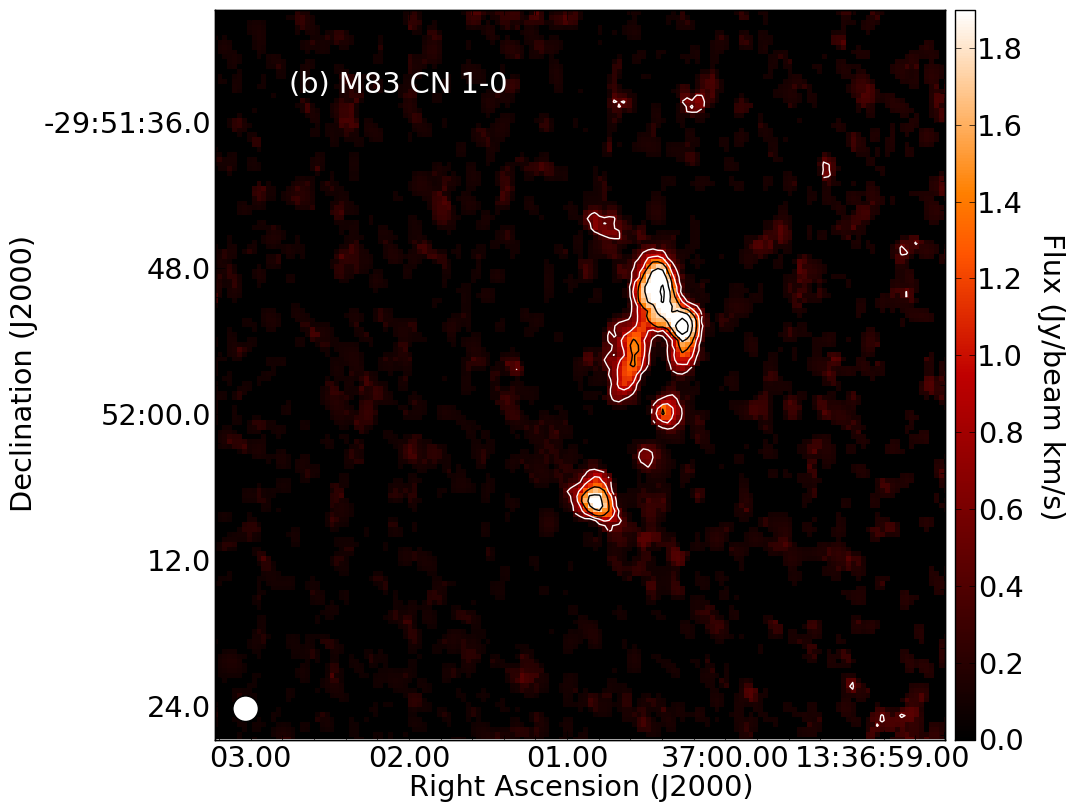}
	\includegraphics[width=\columnwidth]{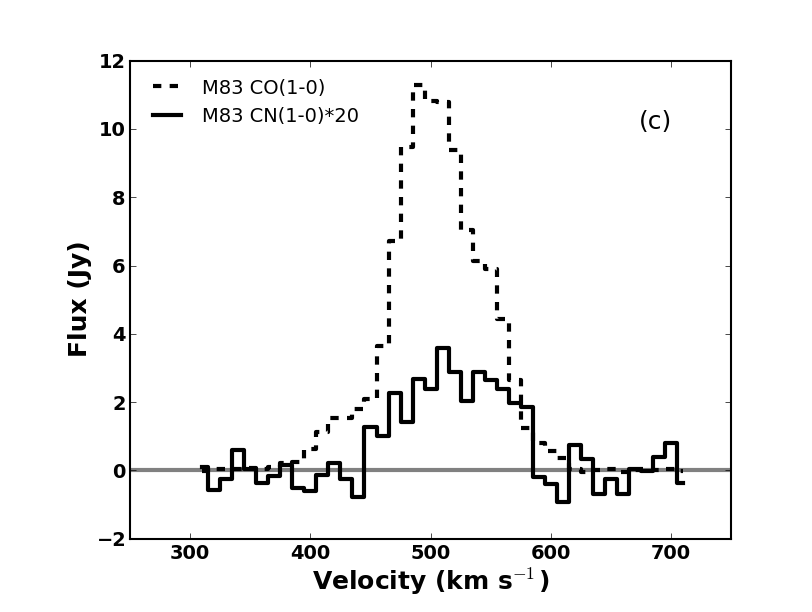}
	\includegraphics[width=\columnwidth]{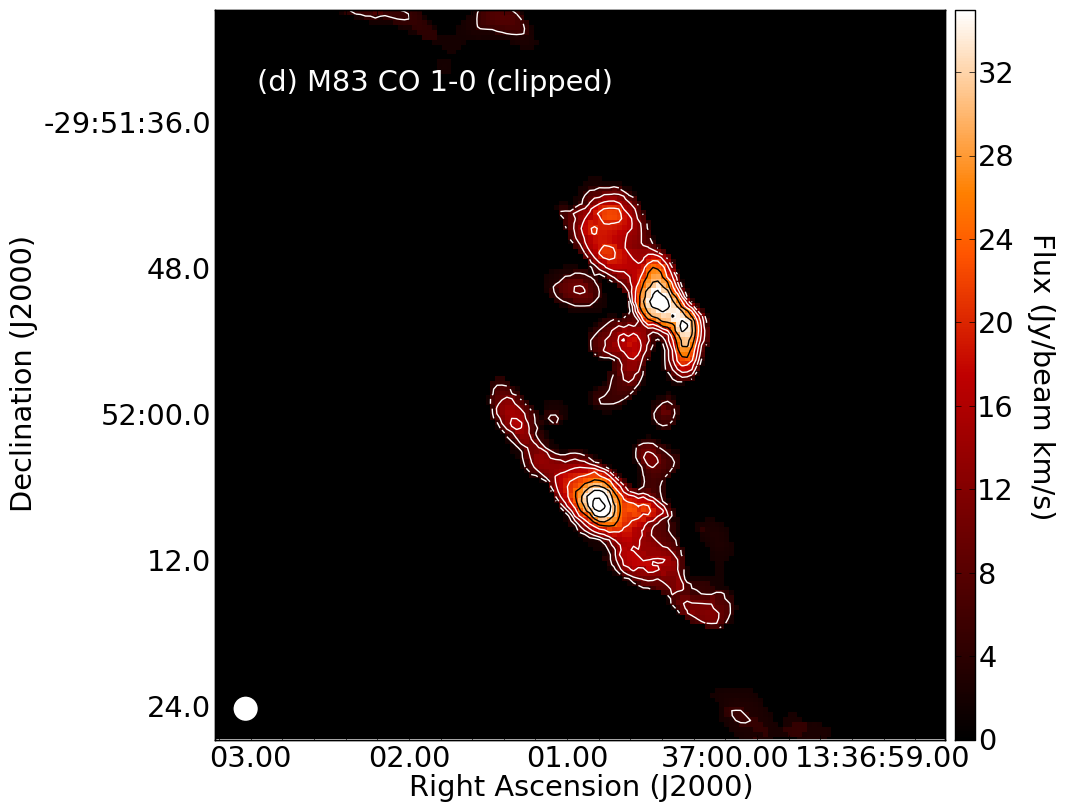}
    \caption{CO $J=1-0$ and CN $N=1-0$ data for M83: (a) CO
      integrated intensity image. Contours are 5, 10, 15 ... 40 Jy
      beam$^{-1}$ km s$^{-1}$. The beam is shown by the ellipse in the
      lower left corner. (b) CN
      integrated intensity image. Contours are 0.45, 0.9, 1.35, 1.8, 2.25 Jy
      beam$^{-1}$ km s$^{-1}$. (c) CO and CN spectra
      integrated over the entire emission region for each line. The CN
    spectrum has been multiplied by a factor of 20. (d) CO integrated intensity image
      clipped to match the S/N of the CO image; contours are the same
      is in panel (a). }
    \label{fig-m83}
\end{figure*}

M83 is a nearby spiral galaxy hosting a central starburst. The CO and
CN emission in this galaxy is highly resolved
and there is a factor of $\sim 3$ variation in the CO/CN ratio.
More investigation into the
individual cloud properties in this galaxy is left to a future paper.
Images and global spectra for M83 are shown in
Fig.~\ref{fig-m83} while the image of the CO/CN line ratio is shown in
Fig.~\ref{fig-ngc253}. Wider-field images of the CO emission in M83
show that this central region has higher CO peak
temperatures and larger line widths than in the bar or spiral arm region
\citep{egusa2018}.

\subsection{NGC 253} 

\begin{figure*}
	\includegraphics[width=\columnwidth]{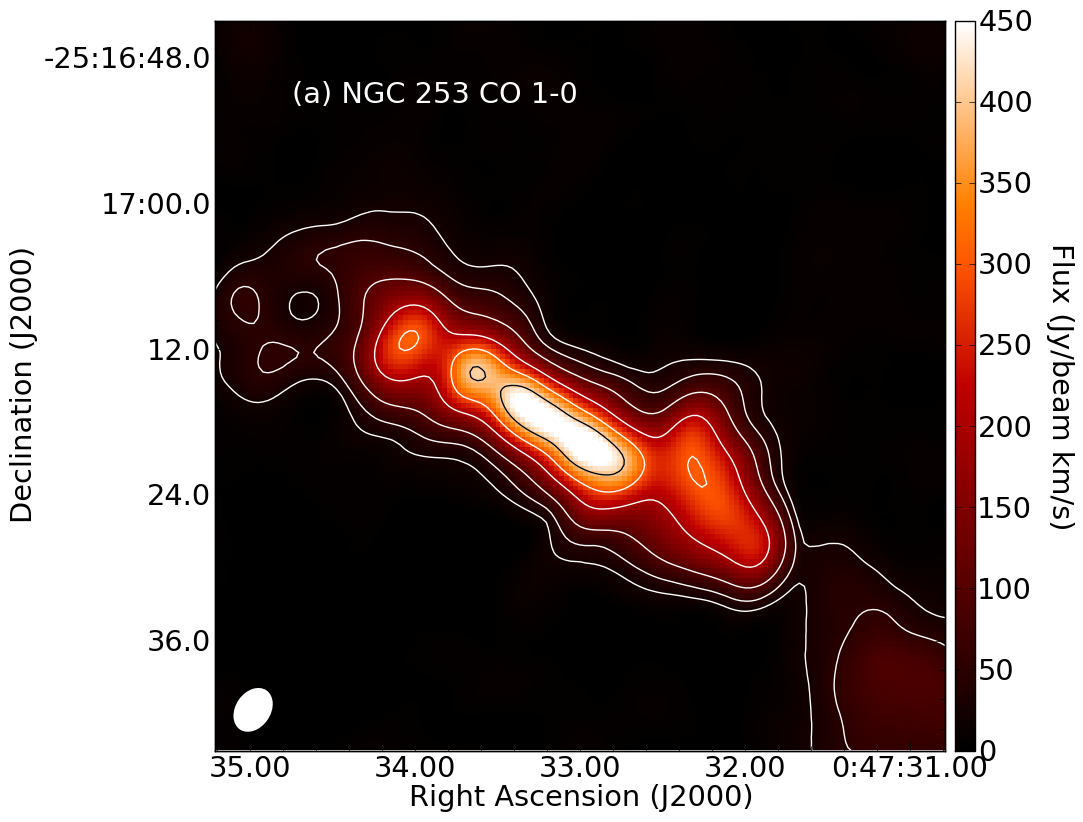}
	\includegraphics[width=\columnwidth]{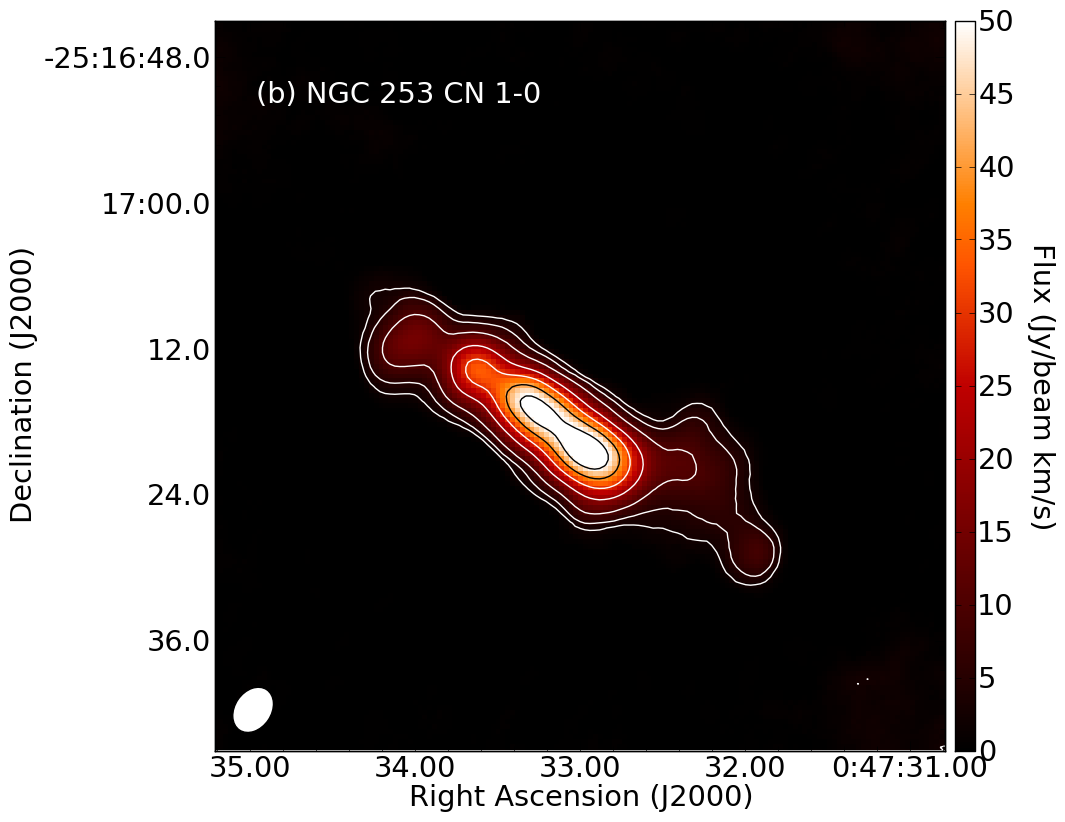}
	\includegraphics[width=\columnwidth]{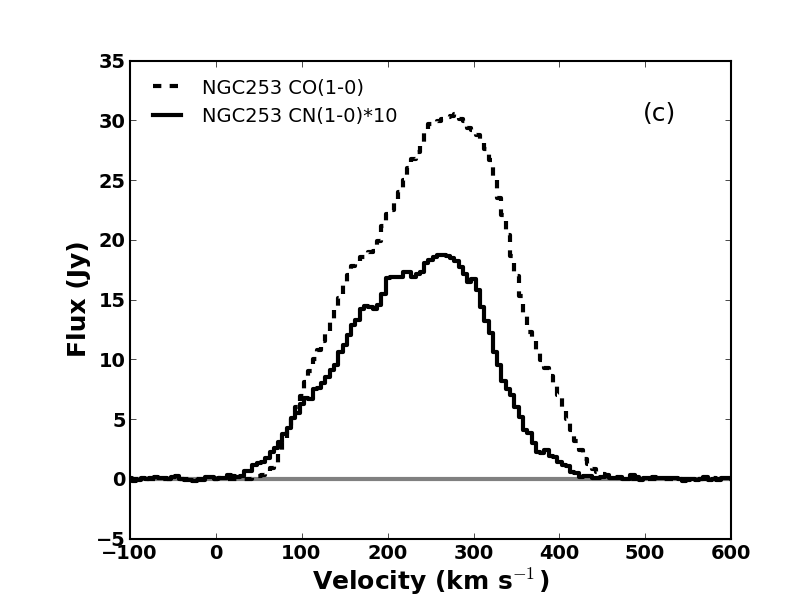}
	\includegraphics[width=\columnwidth]{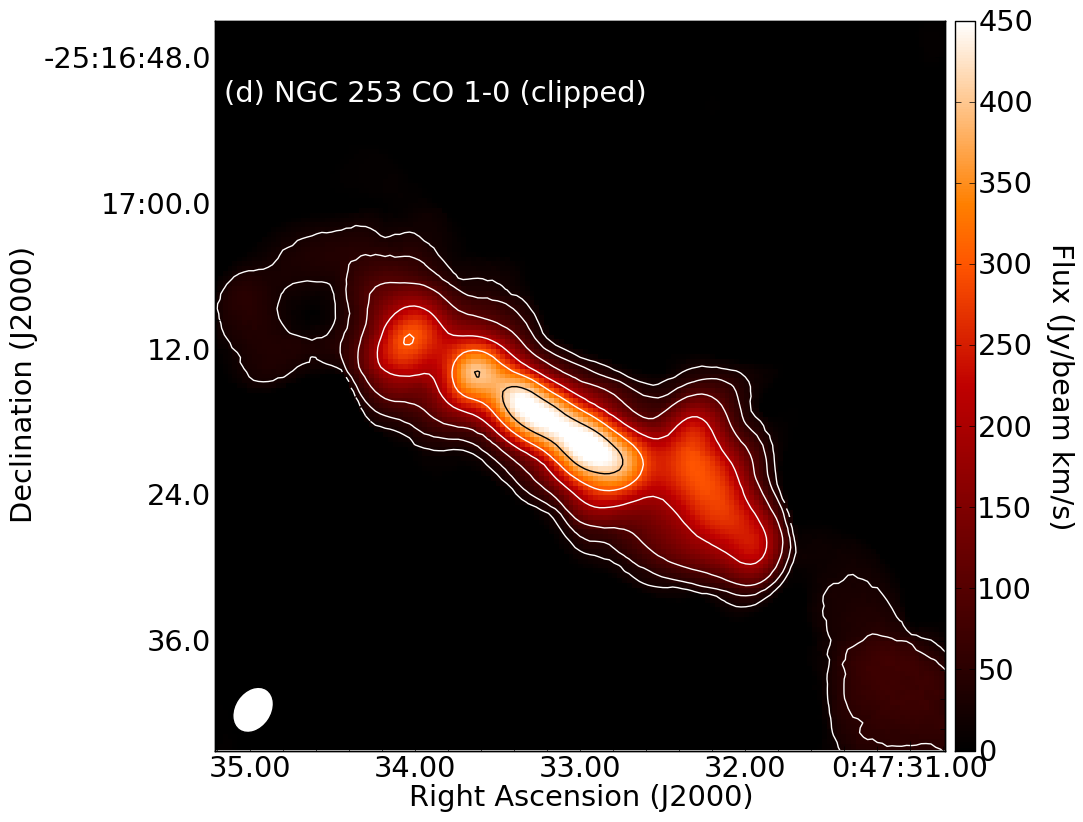}
    \caption{CO $J=1-0$ and CN $N=1-0$ data for NGC253: (a) CO
      integrated intensity image. Contours are 25, 50, 100, 200, 300,
      400, 500 Jy
      beam$^{-1}$ km s$^{-1}$. The beam is shown by the ellipse in the
      lower left corner. (b) CN
      integrated intensity image. Contours are 2.5, 5, 10, 20, 30, 40 50 Jy
      beam$^{-1}$ km s$^{-1}$. (c) CO and CN spectra
      integrated over the entire emission region for each line. The CN
    spectrum has been multiplied by a factor of 10. (d) CO integrated intensity image
      clipped to match the S/N of the CO image; contours are the same
      is in panel (a). }
    \label{fig-ngc253images}
\end{figure*}

NGC 253 is an edge-on spiral galaxy with a strong central starburst.
It has a starburst-driven
outflow in which dense gas tracers such as HCN and CN are detected
\citep{walter2017}. The CO/HCN line ratio is $\sim 10$ in the outflow
and central starburst but falls to $\sim 30$ in the disk
\citep{walter2017}. Individual giant molecular clouds appear as peaks
in dense gas tracers as well as 
in the CO/HCN ratio map at 35 pc resolution
\citep{leroy2015}. \citet{ando2017} observe significant variations in
the spectra of the densest clumps at 10 pc resolution, with the most
line-rich clump resembling a giant hot molecular core.
Images and global spectra for NGC 253 are given in
Fig.~\ref{fig-ngc253images} while the image of the CO/CN line ratio is
shown in Fig.~\ref{fig-ngc253}. 

\subsection{NGC 1377} 

\begin{figure*}
	\includegraphics[width=\columnwidth]{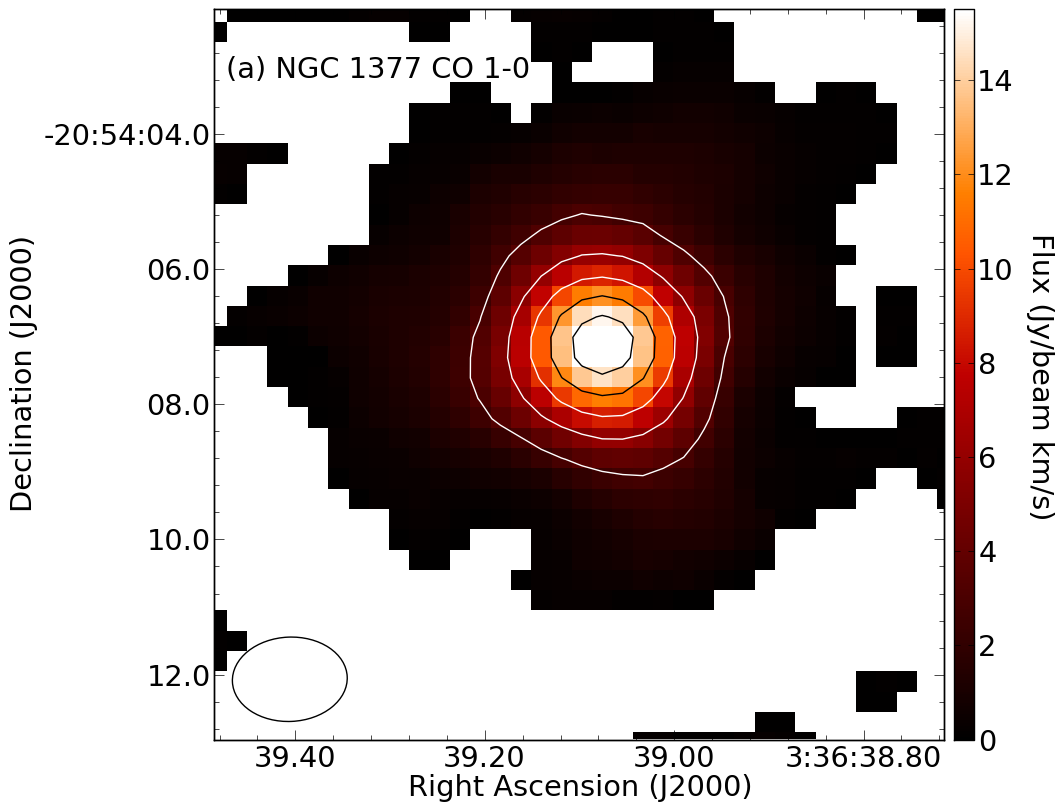}
	\includegraphics[width=\columnwidth]{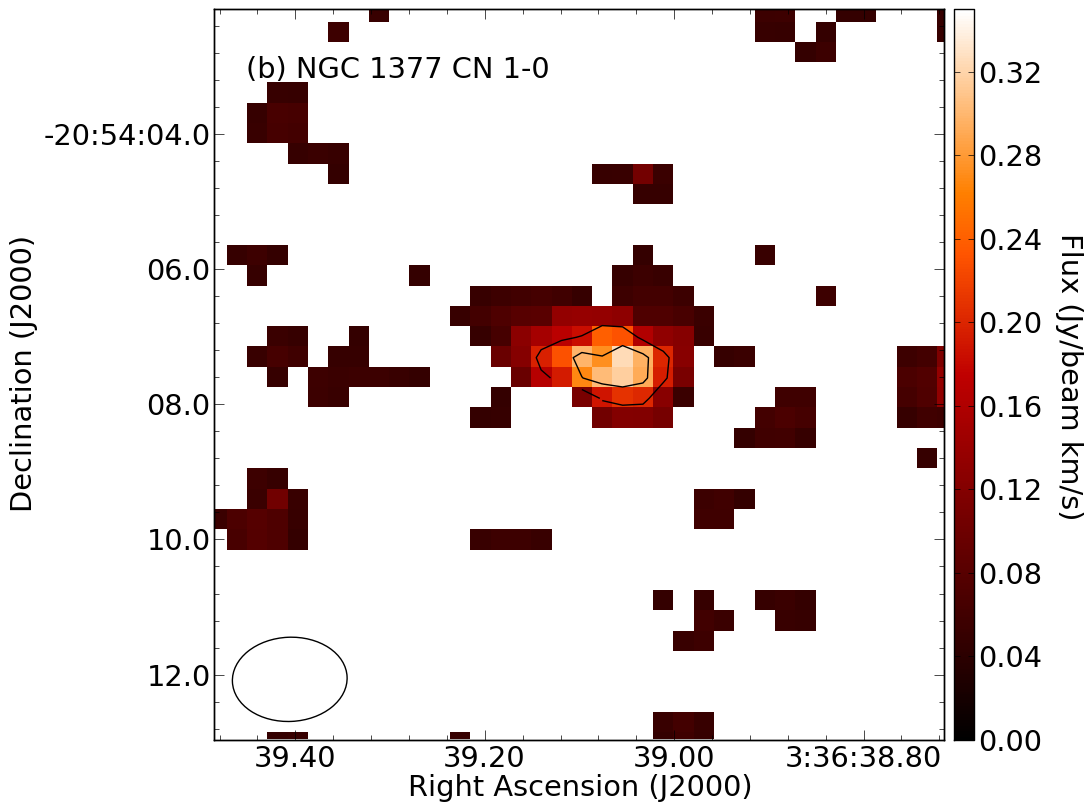}
	\includegraphics[width=\columnwidth]{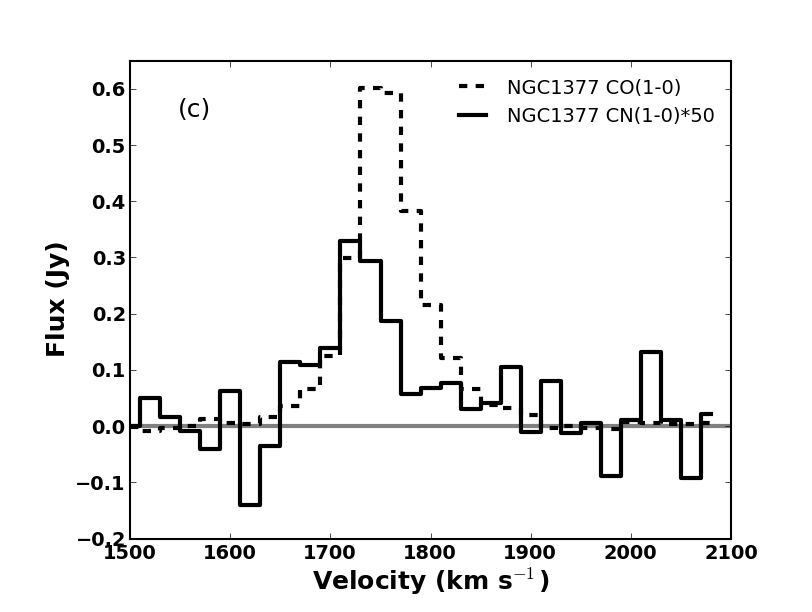}
	\includegraphics[width=\columnwidth]{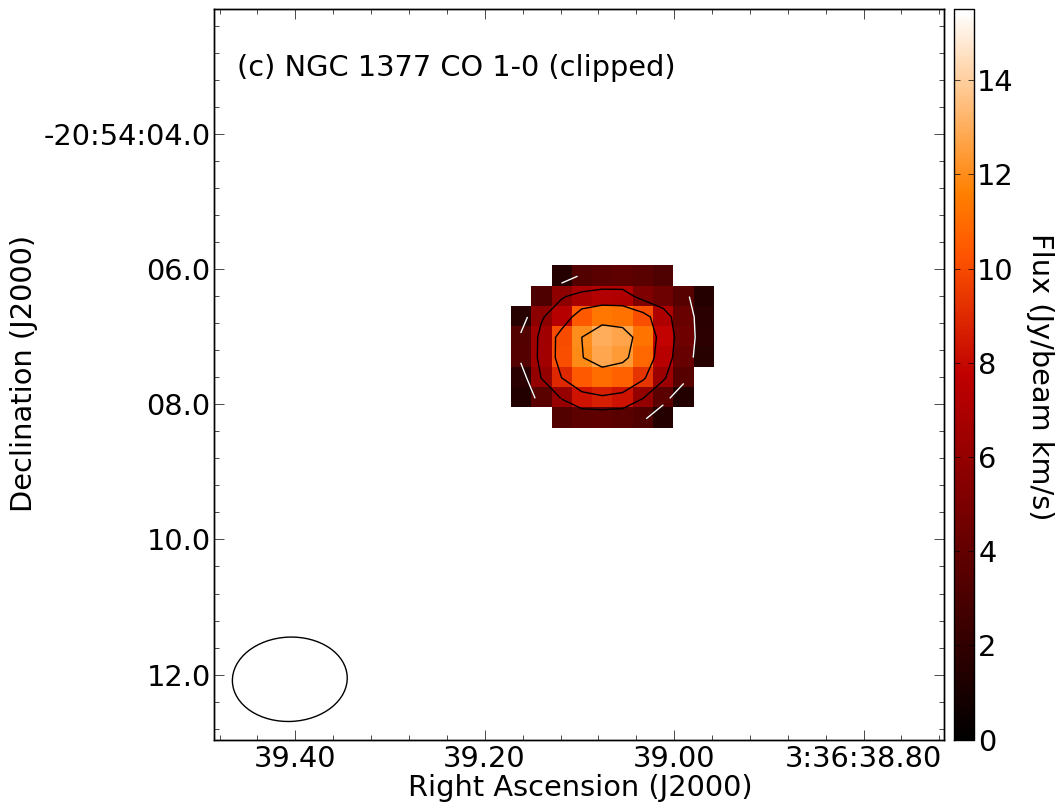}
    \caption{CO $J=1-0$ and CN $N=1-0$ data for NGC 1377: (a) CO
      integrated intensity image. Contours are 3, 6, 9, ... 15 Jy
      beam$^{-1}$ km s$^{-1}$. The beam is shown by the ellipse in the
      lower left corner. (b) CN
      integrated intensity image. Contours are 0.18, 0.27, .36 Jy
      beam$^{-1}$ km s$^{-1}$. (c) CO and CN spectra
      integrated over the entire emission region for each line. The CN
    spectrum has been multiplied by a factor of 50. (d) CO integrated intensity image
      clipped to match the S/N of the CO image; contours are the same
      is in panel (a).  }
    \label{fig-ngc1377}
\end{figure*}

The CO emission in this galaxy extends over $\sim 10$
times the area of the detected CN emission.
NGC 1377 has the highest signal-to-noise measurement of its CO flux
after the three bright mergers/starbursts (NGC 3256, NGC 253,
VV 114). This high S/N ratio suggests that the large 
value of the S/N-matched CO/CN line ratio
observed in this galaxy is real.
 However, because of the large CO/CN line ratio, the S/N on the
CN line is quite low (although CN emission is detected at all the
major CO peaks in the image cube) and so 
the global CO/CN ratio
is likely to be somewhat overestimated.
Images and global spectra for NGC 1377 are shown in
Fig.~\ref{fig-ngc1377}. The peak of the CN emission is offset in
velocity from the peak of the CO emission, which implies there may be
spatially unresolved gradients in the CO/CN line ratio in this galaxy.

\subsection{ NGC 3256} 

The CO/CN line ratio varies substantially across this galaxy. The
minimum value of 8 corresponds to the northern nucleus
\citep{neff2003} while the ratio at the peak of the CO emission in the
southern nucleus is 26. These values have been measured in a single
beam but are constant over areas of a few beams around each nucleus (Fig.~\ref{fig-ngc3256}).
Images and global spectra for NGC 3256 are shown in
Fig.~\ref{fig-ngc3256}. 

\subsection{VV 114} 

VV114 shows the most extreme variations in the CO/CN line ratio and
also has a complicated velocity structure. For the global value
for the S/N-matched CO/CN line ratio, the processing parametres are
as described in Table~\ref{table-sample}. However, because of the
large variations, it was impossible to match the CO moment 0
multiplier closely to the global ratio value.

Separate ratio maps were made for the eastern nucleus and overlap
(western) region 
to try to capture this spatial variation. For the eastern
nucleus, the 
ratio map was made using a CO cutoff a factor of 30 times the CN
cutoff and using a velocity range of 5858-6008 km s$^{-1}$. For the
overlap region, the CO cutoff was 60 and the velocity range was
5708-5828 km s$^{-1}$. The resulting ratio maps are shown in
Fig.~\ref{fig-vv114}. Averaged over a single beam, the CO/CN ratio is
28 towards the eastern nucleus and 61 towards the overlap region.

\begin{figure*}
	\includegraphics[width=\columnwidth]{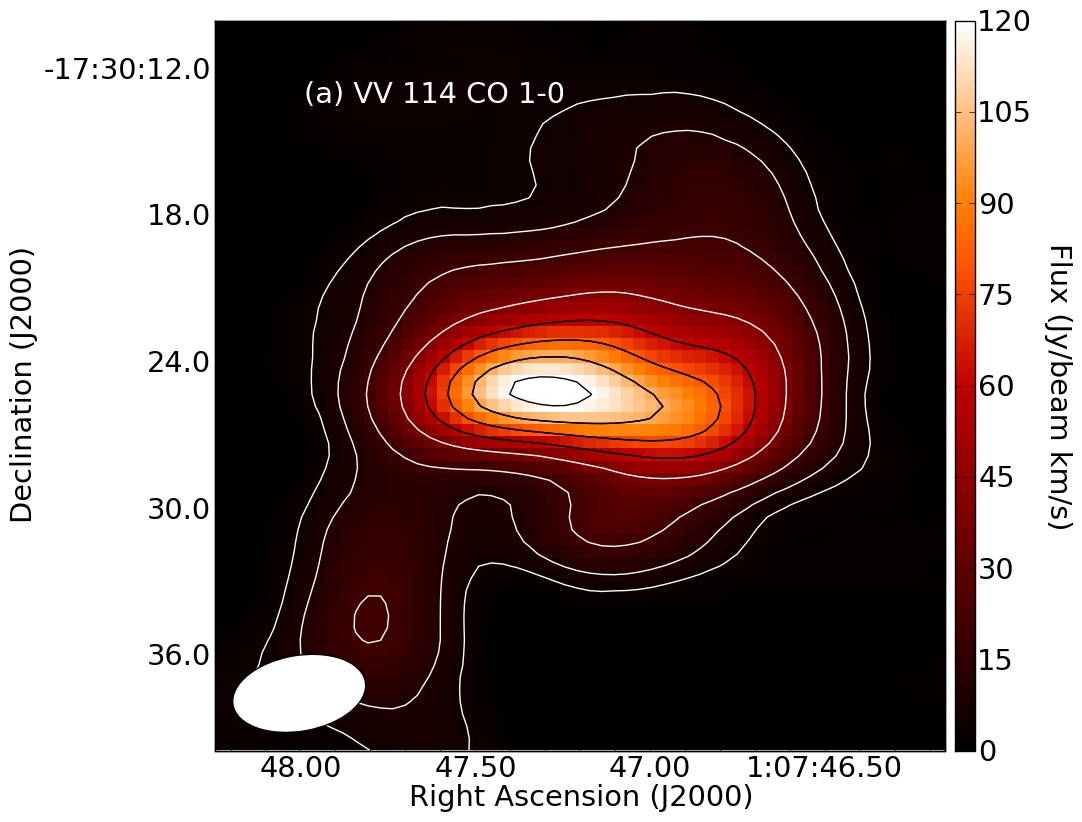}
	\includegraphics[width=\columnwidth]{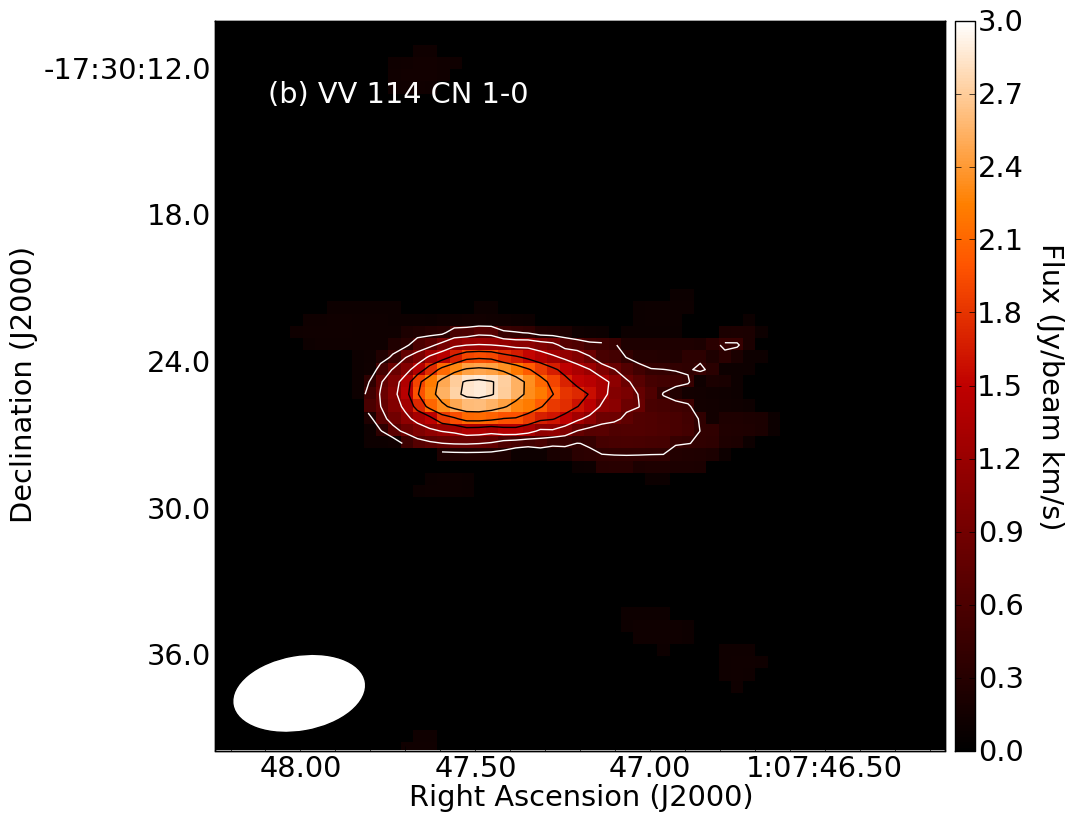}
	\includegraphics[width=\columnwidth]{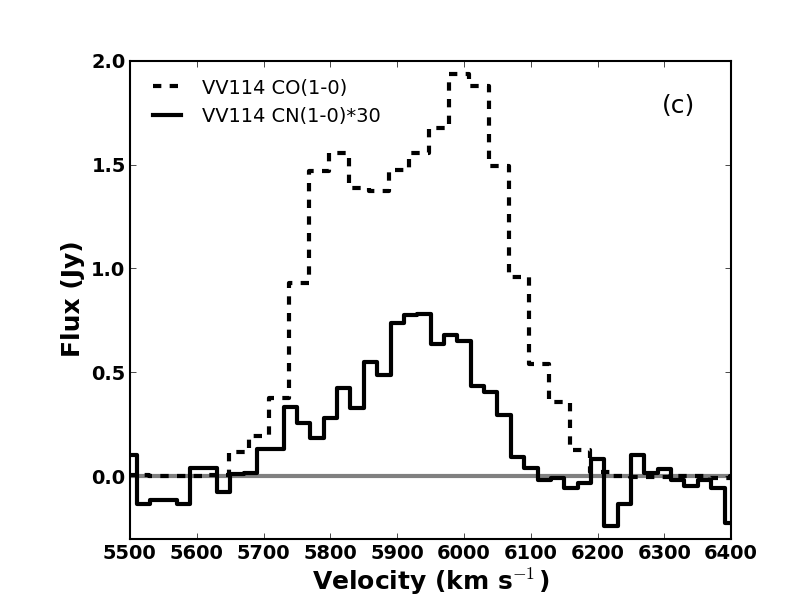}
	\includegraphics[width=\columnwidth]{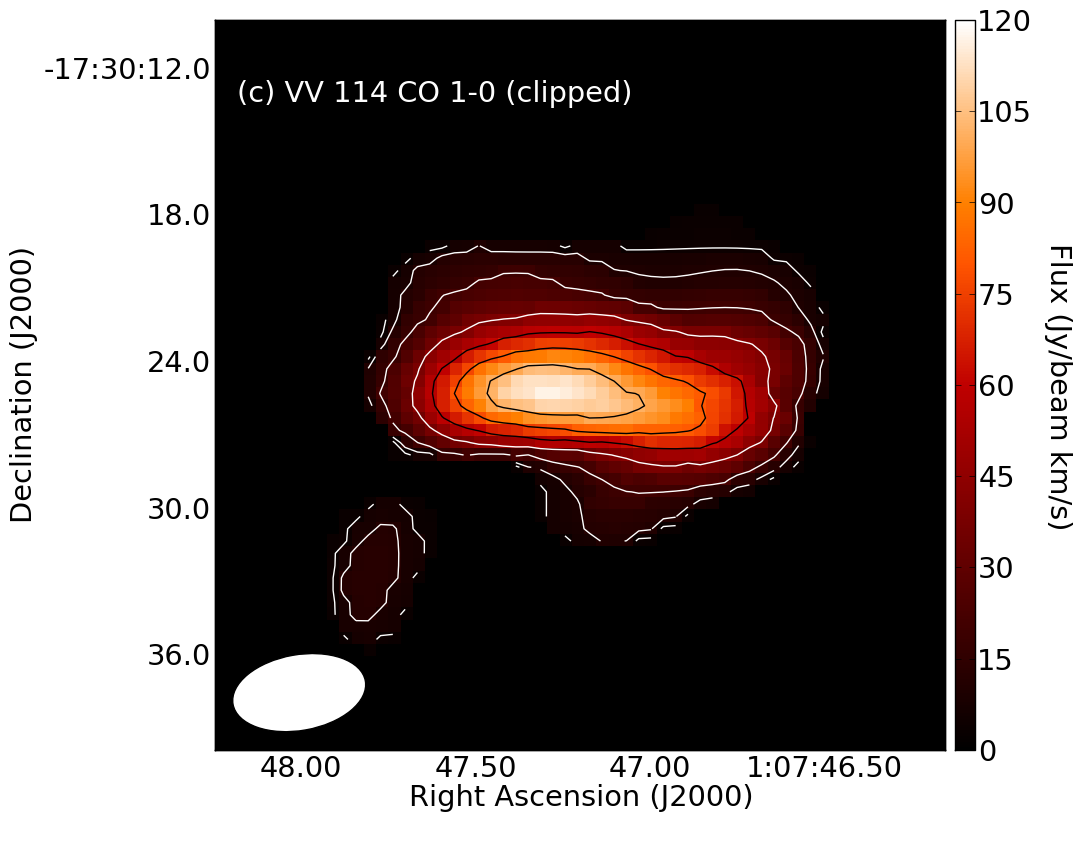}
    \caption{CO $J=1-0$ and CN $N=1-0$ data for VV 114: (a) CO
      integrated intensity image. Contours are 5, 10, 20, 40 ... 120 Jy
      beam$^{-1}$ km s$^{-1}$. The beam is shown by the ellipse in the
      lower left corner. (b) CN
      integrated intensity image. Contours are 0.4, 0.8, 1.2, ... 2.8 Jy
      beam$^{-1}$ km s$^{-1}$. (c) CO and CN spectra
      integrated over the entire emission region for each line. The CN
    spectrum has been multiplied by a factor of 30. (d) CO integrated intensity image
      clipped to match the S/N of the CO image; contours are the same
      is in panel (a). }
    \label{fig-vv114images}
\end{figure*}

The weaker of the two CN line groups was not detected in the overlap
region by \citet{saito2015} and
the upper limit to the emission results in larger CN line ratios 
than would be expected from gas in LTE. These large
line ratios could indicate subthermal excitation of the CN
lines. We re-imaged the emission from VV114 for both CN lines using
natural weighting and 40 km s$^{-1}$ channels; a spectrum (not shown)
towards the 
position of the CO/CN maximum in Fig.~\ref{fig-vv114} suggests
a marginal detection of the weaker CN line, with a ratio or upper
limit consistent with LTE. Thus, we feel there is no need to invoke
sub-thermal excitation to explain the CN emission in the overlap
region of VV 114.
Images and global spectra for VV 114 are shown in
Fig.~\ref{fig-vv114images} 
while images of the CO/CN line ratios in the eastern nucleus and the overlap
region are shown in Fig.~\ref{fig-vv114}.

\section{Lower limits to the CO/CN ratio in 9 galaxies}
\label{app-upperlimits}

We measure lower limits to the CO/CN ratio for 9 galaxies where the CO
emission is  detected at $> 5\sigma$ in
\citet{ueda2014}. 
We adopt a 2$\sigma$ limit on the CN emission and
calculate the lower limits  using two different methods. The first
method combines the peak CO 
emission in a single 20 km s$^{-1}$ channel with the CN rms noise
level measured at the same velocity resolution. This CO/CN line ratio is
most directly comparable to the S/N-matched CO/CN line ratios given in
Table~\ref{table-ratios}. 

For the second method, we made a moment 0
map of the CO emission using all channels in which CO was detected and
measured the peak CO emission from the moment map. We estimated the
corresponding CN integrated intensity rms noise as $20
\sigma_{chan}\sqrt N_{chan}$, where $\sigma_{chan}$ is the rms noise
in the CN data cube, $N_{chan}$ is the number of velocity channels
used to make the CO moment 0 map, and 20 km $^{-1}$ is the velocity
width of the channels in the data cube. These CO/CN line ratios are most
directly comparable to the global CO/CN line ratios given in
Table~\ref{table-ratios}. 

NGC 7252 is detected in CN at the $4\sigma$ level in a single 20 km
s$^{-1}$ channel. The CO/CN ratio derived in this single channel is
consistent with many of the stronger detections in
Table~\ref{table-ratios}. This galaxy is a well-known merger remnant
sometimes known as the ``Atoms for Peace'' galaxy and is rich in star
clusters \citep{miller1997}. The ALMA CO data show a
rotating disk of molecular gas \citep{ueda2014}.

Most of the upper limits in Table~\ref{table-upperlimits} are not
particularly interesting when compared with the CO/CN line ratios in
Table~\ref{table-ratios}. The one possible exception is 
AM 0956-282: the limit CO/CN $> 39$ based on the CO peak channel is
larger than all our detected line ratios except for VV 114 and NGC 1377.
AM 0956-282 is a nearby blue compact dwarf (BCD) galaxy that 
\citet{kim2017} suggest has
recently undergone a flyby interaction with another BCD galaxy. It is
quite nearby ($D_L = 18.6$ Mpc) and has a far-infrared luminosity of only
$7\times 10^8$ L$_\odot$ \citep{ueda2014}.

\begin{table*}
	\centering
	\caption{CO/CN line ratio upper limits in nearby galaxies}
	\label{table-upperlimits}
	\begin{tabular}{lcccc} 
		\hline
		Galaxy & CO peak & CO/CN in 
                & CO peak in &
                CO/CN from \\
& in channel & channel\footnotemark[1] & moment 0 map & moment 0 map\footnotemark[1] \\
& (mJy beam$^{-1}$) & & (Jy beam$^{-1}$ km s$^{-1}$) \\
		\hline
Arp 230 & 69 & $> 15$ & $3.3 \pm 0.2$ & $>11$ \\
Arp 187 & 47 & $> 10$ & $7.9\pm 0.3$ & $> 15 $ \\
AM 0612-373 & 46 & $> 8$ & $7.8\pm 0.3$ & $> 11$ \\
AM 0956-282 &  170 & $>39$ & $5.8\pm 0.1$ & $>28$ \\
NGC 3597 & 114 & $> 20$ & $8.5\pm 0.3$ & $>18$ \\
AM 1255-430 & 33 & $> 11$ & $7.8\pm 0.2$ & $> 30$ \\
AM 2038-382 & 61 & $> 11$ & $8.0 \pm 0.3$ & $> 15$ \\
NGC 7252 & 172 & $16 \pm 4$ & $10.5\pm 0.2$ & $>27$ \\
NGC 7727 & 42 & $>9$ & $6.7\pm 0.3$ & $>16$ \\
		\hline
	\end{tabular}
\\
\footnotemark[1] CO/CN lower limits calculated assuming a $2\sigma$ limit on the CN
emission. \\
\end{table*}


\bsp	
\label{lastpage}
\end{document}